\documentclass[11pt,a4paper]{article}
\pdfoutput=1
\usepackage{jheppub}

\usepackage{multirow}
\usepackage{amsmath,amssymb,amsfonts,leftidx}
\usepackage{graphicx}
\usepackage{color}
\usepackage{braket}
\usepackage{bbm}
\usepackage{url}
\usepackage{slashed}
\usepackage{subfigure}
\usepackage[usenames,dvipsnames]{xcolor}
\usepackage{float} 
\usepackage{epsfig}
\usepackage{graphicx}
\usepackage{euscript}
\usepackage{color}
\usepackage{slashed}
\usepackage{epstopdf}
\usepackage{mathbbol}
\usepackage{mathrsfs}
\usepackage[normalem]{ulem}
\usepackage{pifont}
\usepackage{bbold}
\usepackage{float}
\allowdisplaybreaks
\usepackage{bbding}

\usepackage{verbatim}
\usepackage{graphicx}
\usepackage{latexsym}

\newcommand{\cR}{{\mathcal R}}

\newcommand{\bN}{\bar{N}}
\newcommand{\bs}{\bar{\sigma}}
\newcommand{\hN}{\hat{N}}
\newcommand{\hs}{\hat{\sigma}}

\newcommand{\be}{\begin{equation}}
\newcommand{\ee}{\end{equation}}
\newcommand{\bea}{\begin{align}}
\newcommand{\eea}{\end{align}}

\newcommand{\gb}{\bar{g}} 
\newcommand{\Db}{\bar{D}}

\newcommand{\p}{\partial}

\newcommand{\cG}{\mathcal{G}} 
 
\newcommand{\cO}{\mathcal{O}} 

\newcommand{\cM}{\mathcal{M}}
\newcommand{\cT}{\mathcal{T}}

\definecolor{darkgreen}{rgb}{0.2,0.6,0}
\definecolor{lightblue}{rgb}{0,0.5,0.8}
\definecolor{lightred}{rgb}{0.8,0.2,0.2}
\definecolor{darkorange}{rgb}{1,0.549,0}
\definecolor{brown}{rgb}{0.609, 0.164, 0.164}

\allowdisplaybreaks

\title{Foliated asymptotically safe gravity \\ in the fluctuation approach}

\author[a]{Frank Saueressig,} \emailAdd{f.saueressig@science.ru.nl}

\author[a]{Jian Wang} \emailAdd{jian.wang@science.ru.nl}

\affiliation[a]{Institute for Mathematics, Astrophysics and Particle Physics (IMAPP),
Radboud University, Heyendaalseweg 135, 6525 AJ Nijmegen, The Netherlands}

\abstract{The gravitational asymptotic safety program envisions a high-energy completion of gravity based on a non-Gaussian renormalization group fixed point. A key step in this program is the transition from Euclidean to Lorentzian signature spacetimes. One way to address this challenge is to formulate the quantum theory based on the Arnowitt-Deser-Misner decomposition of the metric field. This equips the Euclidean spacetime with a preferred direction which may serve as the time-direction in the Lorentzian setting. In this work we use the Wetterich equation in order to compute the renormalization group flow of the graviton two-point function. The resulting beta functions possess a non-Gaussian renormalization group fixed point suitable for rendering the theory asymptotically safe. The phase diagram underlying the flow of the two-point function is governed by the interplay between this non-Gaussian fixed point, the Gaussian fixed point, and an infrared fixed point. The latter ensures that the renormalized squared graviton mass cannot take negative values. These results are in qualitative agreement with fluctuation computations carried out in the covariant setting. We take this as non-trivial evidence that the asymptotic safety mechanism remains intact when considering quantum gravity on spacetimes carrying a foliation structure. Technically, our work constitutes the first fluctuation computation carried out within the ADM-framework. Therefore, we also provide a detailed discussion of the conceptual framework, highlighting the elements which differ from fluctuation computations in the covariant setting.
}

\begin{document}
\maketitle
\section{Introduction}
\label{sect.Intro}
Time is an intrinsic building block in our description of nature. General relativity implements this structure by the spacetime metric coming with indefinite signature. Adopting the mostly plus convention, the metric of Minkowski space is $\eta_{\mu\nu} = {\rm diag}(-1,+1,+1,+1)$. This entails that the spacetime geometry $\cM$ possesses a ``preferred direction'' inducing a foliation structure, $\cM = \mathbb{R} \times \Sigma_\tau$. Here $\tau \in \mathbb{R}$ gives the ``time-coordinate'' of an event and $\Sigma_\tau$ are the spatial slices defined by the points in $\cM$ with the same value of $\tau$.  

 The spacetime metric determines the causal order of events. The principle of causality then states that no effect should precede its cause. In general relativity this is realized by the propagation of signals being confined to the lightcone. At the level of a quantum field theory in a fixed Minkowski spacetime causality entails that correlation functions vanish outside the light cone. The concept of causality becomes highly non-trivial once one departs from a classical spacetime and moves into the realm of quantum gravity. In this case, quantum fluctuations in the lightcone structure could, e.g., lead to a violation of causality at microscopic scales \cite{Donoghue:2019ecz}. Our understanding of these effects is far from complete though. Owed to technical reasons, many investigations of quantum gravity effects work in an Euclidean setting, where questions related to the propagation of fields are difficult to assess. Moreover, a generic Euclidean spacetime may not support the additional geometrical structures required to transit from Euclidean to Lorentzian signature settings, so that the analytic continuation only works locally \cite{Baldazzi:2018mtl}. Having a well-defined time-direction even in the Euclidean setting requires a foliation of spacetime into spatial slices which are then welded together to form spacetime. 
  
  Technically, the structures required for the existence of such a foliation can be provided by the Arnowitt-Deser-Misner (ADM)-decomposition of the spacetime metric \cite{Arnowitt:1962hi} (also see \cite{Gourgoulhon:2007ue} for a review),
\be\label{ADMvars}
g_{\mu\nu} \mapsto (N, N_i, \sigma_{ij}) \, . 
\ee
In this case, the metric degrees of freedom are encoded in the lapse function $N$, the shift vector $N_i$, and the metric on the spatial slices $\sigma_{ij}$. The resulting foliation allows to implement the analogue of time in the Euclidean setting. This opens the possibility of an analytic continuation to Lorentzian signature. Moreover, transverse-traceless fluctuations of $\sigma_{ij}$ carry two degrees of freedom (polarizations) which are naturally associated with the two physical degrees of freedom related to the graviton \cite{Arnowitt:1962hi,Bardeen:1980kt,Craps:2014wga}.

As an intermediate step towards a full-fledged theory of quantum gravity formulated at Lorentzian signature, one may study an Euclidean setting implementing the foliation structure of spacetime. In the language of \cite{Isham:1992ms}, this corresponds to a setting in which time is identified before quantizing. Within the gravitational asymptotic safety program, reviewed e.g.\ in \cite{Percacci:2017fkn,Reuter:2019byg,Saueressig:2023irs,Reichert:2020mja,Niedermaier:2006wt,Codello:2008vh,Reuter:2012id,Eichhorn:2018yfc,Pawlowski:2020qer,Bonanno:2020bil}, the construction of renormalization group (RG) flows based on ADM-variables has been developed in \cite{Manrique2011,Rechenberger:2012dt,Contillo:2013fua,Biemans:2016rvp,Biemans:2017zca,Houthoff:2017oam}, also see \cite{Knorr:2018fdu} for an implementation of a foliation structure based on a gauge-fixing construction and \cite{Eichhorn:2019ybe,Knorr:2022mvn} for further discussions.\footnote{For recent steps towards formulating RG flows directly in the Lorentzian setting also see \cite{DAngelo:2022vsh,Banerjee:2022xvi,DAngelo:2023tis}.} Conceptually, this line differs from the covariant setting followed in e.g.\ \cite{Reuter:1996cp,Lauscher:2001ya,Codello:2007bd,Machado:2007ea,Benedetti:2009rx,Dietz:2012ic,Eichhorn:2012va,Dona:2013qba,Christiansen:2015rva,Gies:2016con,Falls:2017lst,Eichhorn:2017eht,Eichhorn:2018akn,Fehre:2021eob,Laporte:2021kyp,Knorr:2021slg,Knorr:2022dsx,Pastor-Gutierrez:2022nki,deBrito:2023myf} in the sense that in the latter case time plays no role at all. This also has profound consequences for the concept of background independence. The construction of the Wetterich equation for gravity \cite{Reuter:1996cp} makes manifest use of the background field formalism. Background independence is then achieved by quantizing the fluctuation field in all backgrounds simultaneously \cite{Becker:2014qya,Dietz:2015owa,Pagani:2019vfm,Becker:2020mjl}, also see \cite{Labus:2016lkh,Mandric:2022dte} for recent developments. The adaptation to the ADM-formalism also builds on the background formalism and follows the same logic. The conceptual difference is that the backgrounds in the ADM-formalism admit a foliation structure by construction. In this sense, the formalism quantizes the fluctuation field in all backgrounds carrying a foliation structure. This guarantees that the background spacetimes possess a Lorentzian analogue. The foliation structure is also an essential element of the Causal Dynamical Triangulations program \cite{Ambjorn:2012jv,Loll:2019rdj}, initiated in \cite{Ambjorn:1998xu,Ambjorn:2000dv}.

 In this work we follow the path initiated in \cite{Manrique:2011jc} and use the Wetterich equation to investigate the RG flow of gravity with the gravitational degrees of freedom carried by the ADM-variables \eqref{ADMvars}. The goal of our study is to identify the non-Gaussian fixed points (NGFPs) which could provide a phenomenologically viable high-energy completion of the theory via the asymptotic safety mechanism. Our work goes beyond the background approximation by studying the flow of the graviton two-point function sourced by the three- and four-point vertices of the theory. It constitutes the first investigation of the asymptotic safety mechanism for foliated spacetimes in a fluctuation computation.

Given the novel type of computation, we provide a detailed exposition on setting up fluctuation computations in the ADM-framework. In particular, we highlight the novel features appearing in the projection onto the vertices of the theory which are absent in the covariant approach. Within this general setting, we perform an explicit study of the RG flow of the graviton two-point function obtained from the Einstein-Hilbert action and its extensions encoding the effect of different speeds of light. As our main result, we identify a NGFP suitable for rendering the construction asymptotically safe. While fluctuation computations projecting the RG flow on different sets of couplings than the background computation, we find that the stability properties of the fixed point in these conceptually different approximations are very similar. We interpret this result as evidence for the robustness of the background computations as well as a first hint on the manifestation of effective universality \cite{Eichhorn:2018akn} in the foliation setting. Moreover, it provides a strong indication that the asymptotic safety mechanism for gravity is robust when transiting from the Euclidean setting to backgrounds carrying a foliation structure.

The remainder of this work is organized as follows. Sects.\ \ref{Sect: FRGonFoli} and \ref{sect.3} review the ADM-decomposition of the metric field and the Wetterich equation formulated on foliated spacetimes, respectively. In particular, Sect.\ \ref{sect.32} gives a general introduction to fluctuation field computations in this setting. The beta functions arising from the foliated Einstein-Hilbert truncation and its non-relativistic two-derivative extensions are derived in Sect.\ \ref{Sec.3.1} and the resulting fixed point structure and phase diagrams are given in Sect.\ \ref{RGanaly}. We conclude with a discussion and outlook in Sect.\ \ref{sect.conc}. Technical details underlying our computation have been relegated to two appendices.


\section{Foliated spacetimes in the ADM-formalism}
\label{Sect: FRGonFoli}
We start by reviewing the  Arnowitt-Deser-Misner (ADM) decomposition of the spacetime metric \cite{Arnowitt:1962hi}. In this case, the gravitational degrees of freedom are encoded in the ADM-fields $(N,N_i,\sigma_{ij})$. Let $\mathcal{M}$ be a $d+1$-dimensional manifold equipped with a Euclidean signature metric $g_{\mu\nu}$ and coordinates $x^\mu$. Here Greek letters $\mu,\nu,\cdots$ denotes spacetime indices taking values from $1$ to $d+1$. We assume that $\cM$ can be foliated by a family of spatial hypersurfaces $\Sigma_\tau$, labeled by the ``time''-parameter $\tau$. Points in the same spatial slice then share the same time-coordinate $\tau$ and are labeled by spatial coordinates $y^i$, $i=1,2,\cdots,d$ on $\Sigma_\tau$. We then perform a change of coordinates 
\be\label{foliationcoordinates}
x^\mu \mapsto (\tau, y^i)
\ee
 and introduce the basis vectors
\begin{equation}\label{basisvects}
t^\alpha \equiv \frac{\partial x^\alpha}{\partial \tau} \bigg |_{y^i} \, ,  \qquad e^\alpha_i \equiv \frac{\partial x^\alpha}{\partial y^i} \bigg |_{\tau}.
\end{equation}
We further define the unit vector $n_\alpha$, normal to the surface $\Sigma_\tau$,
\begin{equation}\label{normalvect}
n_\alpha \equiv N \partial_\alpha \tau \, , \qquad n_\alpha e_i^\alpha=0 \, . 
\end{equation}
Here the lapse function $N(\tau,y^i)$ is introduced as a normalization factor. The vector $t^\alpha$ can then be decomposed into components normal and tangent to the surface $\Sigma_\tau$
\begin{equation}\label{talpha}
t^\alpha =  N n^\alpha + N^i e^\alpha_i \, ,
\end{equation}
where $N^i(\tau,y^i)$ is called the shift vector. 

Next, we apply the change of coordinates \eqref{foliationcoordinates} to the line element 
\begin{equation}\label{dsgamma}
ds^2=g_{\mu\nu} dx^\mu dx^\nu=g_{\mu\nu} (t^\mu d\tau + e^\mu_i dy^i)(t^\nu d\tau + e^\nu_j dy^j) \, . 
\end{equation}
Substituting the decomposition \eqref{talpha} and introducing the induced metric on $\Sigma_\tau$, $\sigma_{ij} \equiv g_{\mu\nu}e^\mu_i e^\nu_j$, the line element can be expressed in terms of the lapse, shift, and $\sigma_{ij}$: 
\begin{equation}\label{dsADM}
	ds^2=(N^2 +\sigma_{ij} N^i N^j ) d\tau^2+2 \sigma_{ij} N^i d\tau dy^j+ \sigma_{ij} dy^i dy^j.
\end{equation}
Comparing eqs.\ \eqref{dsgamma} and \eqref{dsADM} then allows to express the spacetime metric in terms of the ADM-fields,
\begin{equation}\label{gtoadm}
g_{\mu\nu }=
\begin{pmatrix}
N^2+N^i N_i  ~~~& N_j\\
N_i~~~ & \sigma_{ij}\\
\end{pmatrix} \, , \qquad 
g^{\mu\nu }=
\begin{pmatrix}
\frac{1}{N^2}~~~&-\frac{ N^j}{N^2}\\
-\frac{N^i}{N^2}~~~ & \sigma^{ij}+\frac{N^i N^j}{N^2}\\
\end{pmatrix}
,
\end{equation}
where the spatial indices $i,j$ are raised and lowered with the induced metric $\sigma_{ij}$.

Next we consider the transformation properties of the spacetime metric $g_{\mu\nu}$ under the full diffeomorphism group. For a general $d+1$-dimensional infinitesimal coordinate transformation $v^\mu(\tau, y^i)$,  the metric transforms as 
\begin{equation}\label{eq2.8}
\delta g_{\mu\nu} = \mathcal{L}_v g_{\mu\nu},
\end{equation}
where $\mathcal{L}_v$ is the Lie derivative of the metric with respect to the  vector $v^\mu $. Decomposing the infinitesimal coordinate transformation into spatial and time-part, 
\be\label{vdec}
v^\mu(\tau, y^i) = (f(\tau,y^i),\xi^i(\tau,y^i)) \, ,
\ee
 eq.\ \eqref{eq2.8} induces the following transformation of the ADM-fields 
\be\label{Ntrafo}
\begin{split}
\delta N  =& \partial_\tau (f N) + \xi^k \partial_k N- N N^i \partial_i f, \\
\delta N_i =& \partial_\tau (N_i f) + \xi^k \partial_k N_i +N_k \partial_i \xi^k +\sigma_{ki}\partial_\tau \xi^k +N_k N^k \partial_i f +N^2 \partial_i f, \\
\delta \sigma_{ij} =& f \partial_\tau \sigma_{ij} +\xi^k \partial_k \sigma_{ij}+\sigma_{jk}\partial_i \xi^k+\sigma_{ik}\partial_j \xi^k +N_j  \partial_i f+N_i \partial_j f. 
\end{split}
\ee
We also give the transformation for $N^i$ for completeness
\begin{equation}
\delta N^i = \partial_\tau (N^i f) + \xi^k \partial_k N^i -N_k \partial_k \xi^i +\partial_\tau \xi^i -N^i N^j \partial_j f +N^2 \sigma^{ij} \partial_j f.
\end{equation}
The full diffeomorphism group contains an important subgroup, the foliation preserving diffeomorphisms. Their action is obtained by restricting the function $f(\tau,y^i)$ to functions of $\tau$ only.

Finally, we are interested in constructing an action describing the dynamics of the spatial metric $\sigma_{ij}$. At this point, we restrict ourselves to interactions containing at most two derivatives with respect to the spacetime coordinates. While this property is not invariant under the RG flow which inevitably generates higher-order derivative interactions, the resulting action serves as a starting point for generating the tensor structures which will be tracked in the RG computation later on.

Since the foliation equips the manifold with a preferred time-direction, it is natural to construct the action from the view of non-relativistic theories where we have the kinetic term first and subsequently add a potential. To describe how the spatial metric $\sigma_{ij}$ changes between different spatial surfaces $\Sigma_\tau$, we introduce the extrinsic curvature
\begin{equation}
K_{ij} \equiv \frac{1}{2} \mathcal{L}_n \sigma_{ij} = \frac{1}{2} N^{-1} (\partial_\tau \sigma_{ij} - D_i N_j - D_j N_i) \, . 
\end{equation}  
Here  $\mathcal{L}_n$ is the Lie derivative of spatial metric with respect to the normal vector $n^\alpha$ and $D_i$ is the covariant derivative on the spatial slice, carrying the Levi-Civita connection constructed from $\sigma_{ij}$. Since the extrinsic curvature $K_{ij}$ also measures the rate of change of the spatial metric along the time direction, one can construct the kinetic term in terms of $K_{ij}$, 
\begin{equation}\label{kineticaction}
S^{K} = \frac{1}{16 \pi G} \int d\tau d^{d} y N \sqrt{\sigma} (\alpha_1 \, K^{ij}K_{ij}-\alpha_2 K^2) \, , 
\end{equation}
with $K=K_{ij} \sigma^{ij}$ being the trace of the extrinsic curvature. Here the coupling $G$ will turn out to be Newton's constant and $\alpha_1$ and $\alpha_2$ are parameters giving the relative weight between the two kinetic terms.

The potential contains all terms which are independent of time derivatives and compatible with the symmetries we want to impose. Insisting on diffeomorphism invariance on the spatial slice and at most two spatial derivatives, this limits the construction to the volume element and the integrated spatial curvature $\leftidx{^{(d)}} R$. Thus,
\be\label{Spot}
 S^V= \frac{1}{16 \pi G} \int d\tau d^{d} y N \sqrt{\sigma} (-\leftidx{^{(d)}} R + 2\Lambda) \, ,
 \ee
 with $\Lambda$ denoting the cosmological constant. Combining eqs.\ \eqref{kineticaction} we arrive at
\begin{equation}\label{actionansatz}
S= \frac{1}{16 \pi G} \int d\tau d^{d} y N \sqrt{\sigma}{ (\alpha_1 K^{ij}K_{ij}-\alpha_2 K^2-\leftidx{^{(d)}} R + 2\Lambda)}.
\end{equation}

Since we started from the non-relativistic theory, all terms in the kinetic and potential terms have their own independent couplings, and we denote the couplings for the kinetic terms as $\alpha_1$ and $ \alpha_2$. The couplings for the spatial curvature and volume element are given by the Newton coupling and cosmological constant respectively, in order to keep consistency with other works. The parameters $\alpha_1$ and $\alpha_2$ introduce a relative scaling between time and spatial directions.\footnote{This is reminiscent of CDT \cite{Ambjorn2005, Ambjorn:2010hu} where the relative scaling of fundamental (squared) length of space and time also appears.} For $\alpha_1 = \alpha_2 = 1$, the action (\ref{actionansatz}) is the Einstein-Hilbert action, and one recovers invariance under the full diffeomorphism group. In general, the action \eqref{actionansatz} is invariant under foliation preserving diffeomorphisms only. 

We also remark that from a physics perspective, the parameterization \eqref{actionansatz} is redundant in the sense that one of the couplings (canonically $\alpha_1$) can be fixed to $\alpha_1 = 1$ by a rescaling of the lapse $N$ followed by a redefinition of the other couplings. The second parameter in the kinetic part has a physical meaning though. It captures a difference in the propagation speed for the trace- and transverse-traceless modes of the spatial metric fluctuation \cite{Loll:2014xja}. At this point, we keep couplings for all operators contained in \eqref{actionansatz} though.

\section{The Wetterich equation on foliated spacetimes}
\label{sect.3}
The Wetterich equation \cite{Wetterich:1992yh,Morris1994,Reuter:1993kw,Reuter:1996cp} encodes the change of the effective average action $\Gamma_k$ when integrating out quantum fluctuations with momenta close to the coarse-graining scale $k$ \cite{Pawlowski:2005xe,Gies2012,Percacci:2017fkn,Reuter:2019byg,Bonanno:2020bil,Dupuis:2020fhh,Saueressig:2023irs}. In this way, it realizes the Wilsonian picture of renormalization. Its adaptation to the ADM-decomposition has been made in \cite{Manrique2011} with further details given in \cite{Rechenberger:2012dt}. In this section, we review the key ingredients of the construction and explain the setup for performing computations tracking the fluctuation fields.
\subsection{General background on the functional renormalization group}
\label{sect.30}
The most-frequently used tool for calculating RG flows in theories containing gravitational degrees of freedom is the the Wetterich equation \cite{Wetterich:1992yh,Morris1994,Reuter:1993kw,Reuter:1996cp}. This equation captures the dependence of the effective average action $\Gamma_k$ on the coarse-graining scale $k$. It realizes the Wilsonian idea of renormalization in the sense that it captures the change of the effective description of the system when integrating out quantum fluctuations with momenta $p^2 \simeq k^2$. The equation takes a one-loop form and is given by  \cite{Wetterich:1992yh,Morris1994}
\begin{equation}\label{WetterichEq}
	\partial_t \Gamma_k = \frac{1}{2} \text{STr} \left[ \left( \Gamma^{(2)}_k + \mathcal{R}_k \right)^{-1} \, \partial_t \mathcal{R}_k \right]\, . 
\end{equation}
Here $t\equiv\text{ ln } ( k / k_0)$ denotes the RG time with $k_0$ being an arbitrary reference scale and $\Gamma^{(2)}_k$ is the second functional derivative of $\Gamma_k$ with respect to the fluctuation fields. The regulator $\cR_k(p^2)$ equips the fluctuations with momenta $p^2 \ll k^2$ with a $k$-dependent mass term and vanishes for momenta $p^2 \gg k^2$. The latter property ensures that the trace on the right-hand side is free from UV-divergences since its argument vanishes sufficiently fast for high momenta. Both $\Gamma^{(2)}_k$ and $\cR_k$ are matrix-valued in field space. The supertrace STr then includes an integration over loop-momenta as well as sums over all fluctuation fields and internal indices. Moreover it provides a minus-sign to the contribution of the ghost fields. 

In practice one obtains non-perturbative, approximate solutions of \eqref{WetterichEq} by projecting the exact equation to a subspace of all admissible action functionals $\cO_i$ constructable from a given field content,
\be\label{truncansatz} 
\Gamma_k \simeq \sum_i \bar{u}^i(k) \, \cO_i \, . 
\ee
The $k$-dependence of $\Gamma_k$ is then captured by the dimensionful couplings $\bar{u}^i(k)$. When analyzing the RG flow it is convenient to consider the dimensionless versions of these couplings constructed with respect to the coarse-graining scale $k$,  $u^i(k) = \bar{u}^i(k) k^{-[d_i]}$, where $[d_i]$ is the mass-dimension of $\bar{u}^i$. Substituting \eqref{truncansatz} into \eqref{WetterichEq} and matching the coefficients multiplying the functionals $\cO_i$ on its left and right-hand side gives the beta functions encoding the flow of the couplings
\be\label{defbeta}
\p_t u^i(k) = \beta_{u^i}(u^j(k)) \, . 
\ee  
The solutions of this system are called RG trajectories.

The most important properties of the beta functions are their fixed points $(u^i_*)$ where
\be\label{fixedpointdef}
 \beta_{u^i}(u^j_*) = 0 \, , \qquad \forall \, i \, . 
 \ee
 Depending on whether the fixed point action corresponds to a free theory or admits interactions, one distinguishes between a Gaussian fixed point (GFP) and a non-Gaussian fixed point (NGFP). 
 
 In the vicinity of a fixed point, the properties of the RG flow can be obtained by linearizing the system \eqref{defbeta} 
 \be\label{stabmat}
 \p_t u^i(k) = \sum_j B^i{}_j (u^j - u^j_*) + O((u^j - u^j_*)^2) \, , \quad 
 B^i{}_j = \left. \frac{\p \beta_{u^i}}{\p u^j} \right|_{u^i = u^i_*} \, .  
 \ee
 The stability properties of the flow are captured by the stability coefficients $\theta_I$, defined as minus the eigenvalues of the stability matrix $ B^i{}_j$. For eigendirections of  $B^i{}_j$ associated with coefficients with Re$\theta_I > 0$ the flow is dragged into the fixed point as $k \rightarrow \infty$ while eigendirections where Re$\theta_I < 0$ are repulsive in this limit.
 
 The asymptotic safety hypothesis then stipulates that the high-energy completion of gravity is provided by a NGFP \cite{Weinberg:1980gg}. If the underlying NGFP has eigendirections where Re$\theta_I < 0$, the presupposition that the NGFP provides the high-energy completion leads to testable predictions in the following sense: any theory meeting this criterion must be situated within the subspace of RG trajectories (the UV-critical hypersurface of the fixed point) emanated from the fixed point as $k$ decreases. This induces conditions among the coupling constants which are testable, at least in principle. Recently, it has been argued in \cite{Pastor-Gutierrez:2022nki}, that applying this condition to the standard model of particle physics allows to identify the interacting gravity-matter fixed point providing the UV-completion of the theory.  
 
\subsection{Introducing the foliation structure}
\label{sect.31}
The construction of the effective average action for theories treating the gravitational degrees of freedom at the quantum level makes manifest use of the background field method \cite{Reuter:1996cp}. The presence of the background structure is essential for dividing the spectrum of fluctuations into long-range and short-range with respect to the coarse-graining scale $k$. In essence, the use of background fields allows to treat quantum fluctuations in the metric along the lines of matter degrees of freedom quantized within the framework of quantum field theory in a curved spacetime. On top of this conceptual need, the background field formalism also induces auxiliary symmetries in $\Gamma_k$ which constrain the type of interaction monomials which are generated along the RG flow. 

In practice, we implement the background field formalism by starting from the ADM-fields $\chi = (N,N_i,\sigma_{ij})$ and decomposing them into a background part $\bar{\chi} = (\bN,\bN_i,\bs_{ij})$ and fluctuations $\hat{\chi} = (\hN,\hN_i,\hs_{ij})$ via a linear split\footnote{Since the ADM-decomposition of the metric \eqref{gtoadm} is non-linear in the ADM-fields, the vertices containing a fixed power of the fluctuation fields capture different contributions than the ones obtained in the covariant computation building on the split $g_{\mu\nu} = \gb_{\mu\nu} + h_{\mu\nu}$.}
\begin{equation}\label{linsplit}
	N= \bar{N}+\hat{N} \, , \qquad N_i =\bar{N}_i +\hat{N}_i \, , \qquad \sigma_{ij} =\bar{\sigma}_{ij} +\hat{\sigma}_{ij}.
\end{equation}

Depending on the values of $\alpha_1$ and $\alpha_2$, the action \eqref{actionansatz} is invariant either with respect to the full diffeomorphism group or foliation preserving diffeomorphisms. In order to obtain well-defined propagators, the gravitational part of the effective average action must thus be supplemented by a gauge-fixing condition and the corresponding ghost contribution. Thus the ADM-adaptation of $\Gamma_k$ has the general structure
\be\label{Gammakgen}
\begin{split}
\Gamma_k = & \, \bar{\Gamma}_k[N,N_i,\sigma_{ij}] + \widehat{\Gamma}_k[\hN,\hN_i,\hs_{ij};\bN,\bN_i,\bs_{ij}] + \Gamma_k^{\rm gf}[\hN,\hN_i,\hs_{ij};\bN,\bN_i,\bs_{ij}] \\ & \, + \Gamma_k^{\rm ghost}[\hN,\hN_i\hs_{ij},\bar{c},\bar{b}^i,c,b_i;\bN,\bN_i,\bs_{ij}] \, . 
\end{split}
\ee
Here $\bar{\Gamma}_k[N,N_i,\sigma_{ij}]$ is the ``diagonal'' part of the action depending on the background and fluctuation fields in the combination \eqref{linsplit} only and $\widehat{\Gamma}_k$ encodes the ``off-diagonal'' contributions and genuinely depends on both arguments. $\Gamma_k^{\rm gf}$ provides the gauge-fixing of the action and is accompanied by the action for the ghost $(c, b^i)$ and anti-ghost fields $(\bar{c},\bar{b}_i)$ capturing the contribution of the Faddeev-Popov determinant. Eq.\ \eqref{Gammakgen} anticipates that all sectors may contain $k$-dependent couplings.

Concretely, we follow \cite{Biemans:2017zca, Barvinsky2016} and work within the class of background gauge-fixings. The main idea underling our choice of gauge fixing is to introduce a contribution bilinear in the fluctuation fields which equips all fields with a relativistic dispersion relation. Restricting the constructions to terms with at most two derivatives, we can parameterize the gauge-fixing functional as  
\begin{equation}\label{Gammagf}
	\Gamma^{\text{gf}}_k = \frac{1}{32 \pi G_k} \int d\tau d^d x \bN \sqrt{\bar{\sigma}} \left[ F_i \bar{\sigma}^{ij} F_j +F^2 \right] \, .
\end{equation}
The functionals $F$ and $F_i$ are linear in the fluctuation fields and implement the gauge-fixing condition. The most general form of $F$ and $F_i$ is a linear combination of the fluctuation fields including one temporal or spatial derivative. In a flat background, their generic form is given by
\begin{equation}
	\begin{split}
		F = & c_1 \, \partial_\tau \hat{N} + c_2 \, \partial^i \hat{N}_i + c_3  \, \partial_\tau \hat{\sigma} ,\\
		F_i = & c_4  \, \partial_\tau \hat{N}_i + c_5 \, \partial_i \hat{N} + c_6 \, \partial_i \hat{\sigma} + c_7 \, \partial^j \hat{\sigma}_{ji}.
	\end{split}
\end{equation}
The free coefficients $c_1,\cdots,c_7$ can be chosen for later convenience. In the sequel, we will fix 
\begin{equation}\label{gaugechoice}
	\begin{split}
		c_1=-\sqrt{2},~ c_2=-\sqrt{2},~c_3=\frac{1}{\sqrt{2}},
		c_4=-\sqrt{2},~ c_5=\sqrt{2},~ c_6=\frac{1}{\sqrt{2}},~ c_7=-\sqrt{2},
	\end{split}
\end{equation}
which implements the harmonic gauge condition at the level of the ADM-decomposition. The ghost action exponentiating the Faddeev-Popov determinant is obtained in the standard way
\begin{equation}\label{ghostgen}
	\begin{split}
		\Gamma^{\text{ghost}}_{k}&=
		 \int d\tau d^d x \bN \sqrt{\bar{\sigma}}  \left[  \bar{c} \, \frac{\delta F}{\delta \hat{\chi}^{a}} \left( \delta_{c, b_j} \chi^{a} \right) + \bar{b}^i \, \frac{\delta F_i}{\delta \hat{\chi}^{a}} \left( \delta_{c, b_j} \chi^{a} \right) \right] \, . 
	\end{split}
\end{equation}
Here $(\delta_{c, b_j} \chi^{a})$ is the transformation of ADM-field $\chi^{a}$ introduced in \eqref{Ntrafo}, with $f$ and $\xi^i$ replaced by $c$ and $b^i$. For the parameters \eqref{gaugechoice}, the evaluation of \eqref{ghostgen} leads to a rather lengthy expression. Its explicit form is given in App.\ \ref{App.D}. 

The dependence of $\Gamma_k$ on the coarse-graining scale $k$ is then governed by the Wetterich equation \eqref{WetterichEq}. At this point a technical remark about the construction of the regulator function $\cR_k(\Box)$ is in order (also see \cite{Codello:2008vh,Braun:2022mgx} for a more detailed discussion). In practice, the regulator is a non-local function of a differential operator $\Box$ which is used to discriminate between fluctuations coming with ``high-'' and ``low-''momentum with respect to the coarse graining scale $k$. Its basic property is that it provides a mass term for low-momentum fluctuation and decays sufficiently fast for the high-momentum modes. In the covariant setting it is natural to use the background Laplacian, $\Box \equiv -\gb^{\mu\nu} \Db_\mu \Db_\nu$, potentially supplemented by an endomorphism constructed from the background curvature, in order to ``measure'' the momentum of a fluctuation \cite{Codello:2008vh}. In the foliated setting, we have a natural discrimination between spatial and ``time''-derivatives. This opens more options. In particular, one can resort to a regularization procedure where $\Box$ does not contain derivatives with respect to the ``time''-direction. The discrimination between low- and high-momentum fluctuations can then be based on the Laplacian constructed on the spatial slices $\Sigma_\tau$, $\Box \equiv - \bs^{ij} \Db_i \Db_j$. This choice still realizes a $k$-dependent mass term for the fluctuation fields and suffices to render the trace on the right-hand side of eq.\ \eqref{WetterichEq} finite. This choice comes with the advantages that it does not induce higher-order time-derivatives in the regularization procedure. Moreover, it orders fluctuations in a positive semi-definite way which can be carried over to Lorentzian signature computations. This is the route taken in the background computations \cite{Manrique:2011jc,Rechenberger:2012dt,Contillo:2013fua,Biemans:2016rvp,Biemans:2017zca}.
As a drawback, the regulator introduces a non-covariant element in the construction which sources diffeomorphism-violating effects in the RG flow. Since our present work is limited to the Euclidean signature setting, we follow the construction \cite{Knorr:2018fdu} and adopt a covariant choice for $\Box$. It turns out that this choice is also technically preferred when carrying out fluctuation computations in the foliated setting.
%

%
\subsection{Solving the Wetterich equation in the fluctuation approach}
\label{sect.32}
Before delving into the actual computation, let us first introduce the conceptual elements underlying fluctuation computations based on the Wetterich equation, following \cite{Pawlowski:2020qer}. The general setup will be introduced in Sect.\ \ref{sect.3.2.1} and we give an instructive example based on the Einstein-Hilbert action in Sect.\ \ref{sect.3.2.2}. Details specific to RG flows on a foliated spacetime are discussed in Sect.\ \ref{sect.3.2.3}. 

\subsubsection{Fluctuation field computations - the general setup}
\label{sect.3.2.1}
We start with briefly reviewing the fluctuation approach for the functional renormalization group. Generically, we consider a theory whose field content comprises $N$ fields collectively denoted by $\chi \equiv (\chi_1,\cdots,\chi_N)$. In the ADM-formalism, $\chi$ contains the lapse, shift, and spatial metric, as well as the ghost and anti-ghost fields. The background field method decomposes these fields into their background parts $\bar{\chi} \equiv (\bar{\chi}_1,\cdots,\bar{\chi}_N)$ and fluctuations $\hat{\chi} \equiv (\hat{\chi}_1,\cdots,\hat{\chi}_N)$. This decomposition could be implemented through the linear split $\chi_a = \bar{\chi}_a + \hat{\chi}_a$, $a=1,\cdots,N$. Typically, the background fields associated with the ghost and anti-ghost fields are taken to be zero.

The idea of the fluctuation approach is to expand the effective average action in powers of the fluctuation fields
\begin{equation}\label{vertexpan}
	\Gamma_k [\hat{\chi};\bar{\chi}]=\sum_{n=0}^\infty P(a_i) \int_x \Gamma^{(\hat{\chi}_{a_1} \cdots  \hat{\chi}_{a_n})}_k [\bar{\chi}]~ \hat{\chi}_{a_1} \cdots  \hat{\chi}_{a_n} \, . 
\end{equation}
The dependence of $\Gamma_k$ on the coarse-graining scale is carried by the expansion coefficients $\Gamma^{(\hat{\chi}_{a_1} \cdots  \hat{\chi}_{a_n})}_k [\bar{\chi}]$. These depend on background quantities only and may contain covariant derivatives acting on the fluctuation fields. The $P(a_i)$ denotes a combinatorial factor ensuring that
\be\label{vertex-def}
\Gamma^{(\hat{\chi}_{a_1} \cdots  \hat{\chi}_{a_n})}_k [\bar{\chi}] = \left. 
\frac{\delta^{m}}{\delta \hat{\chi}_{a_1} \cdots  \delta \hat{\chi}_{a_m}} \Gamma_k [\hat{\chi};\bar{\chi}]
 \right|_{\hat{\chi}=0} \, , 
\ee
and a sum over the indices $a_i$ is implied. The expansion disentangles the contributions from the background and fluctuation fields. In explicit computations, it is useful to extract the wave-function renormalization factors of the fluctuation fields from the vertices, 
\be
\Gamma^{(\hat{\chi}_{a_1} \cdots  \hat{\chi}_{a_n})}_k [\bar{\chi}] = \left(  \prod_{i=1}^n Z^{\frac{1}{2}}_{{\hat{\chi}}_{a_i}} \right) ~ \bar{\Gamma}_k^{({\hat{\chi}}_{a_1} \cdot \cdot \cdot {\hat{\chi}}_{a_n})} [\bar{\chi}] \, . 
\ee
In order to keep our notation light, we will keep these factors within $\Gamma^{(\hat{\chi}_{a_1} \cdots  \hat{\chi}_{a_n})}_k [\bar{\chi}]$.

At this stage, one can introduce a basis on the space of tensor structures $(\mathcal{T}^{({\hat{\chi}}_{a_1} \cdots {\hat{\chi}}_{a_n})})$ contracting $n$ fluctuation fields. In general, these basis elements carry the dependence of the vertices on the internal indices of the fields. The expansion coefficients $\Gamma^{(\hat{\chi}_{a_1} \cdots  \hat{\chi}_{a_n})}_k [\bar{\chi}]$ can then be expanded in this basis
\begin{equation}\label{tensorexp}
	\Gamma_k^{({\phi}_{a_1} \cdot \cdot \cdot {\phi}_{a_n})} [\bar{\chi}] = \sum_j
	\bar{u}_{n,j}(k) \, 
	\mathcal{T}_j ^{({\phi}_{a_1} \cdot \cdot \cdot {\phi}_{a_n})}[\bar{\chi}].
\end{equation}
The expansion coefficients $\bar{u}_{n,j}(k)$ are the dimensionful coupling constants of the theory and depend on the coarse-graining scale $k$.

The dependence of the couplings $\bar{u}_{n,j}(k)$ on $k$ can then be obtained by substituting the expansion \eqref{vertexpan} into the Wetterich equation \eqref{WetterichEq} and comparing the coefficients multiplying a given tensor structure on its left- and right-hand side. The projection onto the tensor structures is conveniently performed by taking functional derivatives of the initial equation with respect to the fluctuation fields and their contracted (generalized) momenta. Making use of \eqref{vertex-def}, this leads to equations of the form
\be\label{npthierarchy}
\p_t \Gamma^{(\hat{\chi}_{a_1} \cdots  \hat{\chi}_{a_n})}_k [\bar{\chi}] = \frac{1}{2}
\left. 
\frac{\delta^{n}}{\delta \hat{\chi}_{a_1} \cdots  \delta \hat{\chi}_{a_n}} {\rm Str}\left[ 
 ( \Gamma^{(2)}_k[\hat{\chi};\bar{\chi}] + \mathcal{R}_k[\bar{\chi}] )^{-1} \, \partial_t \mathcal{R}_k[\bar{\chi}]  
\right]
\right|_{\hat{\chi}=0} \, . 
\ee
For the one-point correlation functions this general expression implies
\begin{eqnarray}
	\partial_t \Gamma^{(\hat{\chi}_i)}_k &=& - \frac{1}{2} \text{STr} \left. \left[ \left(\Gamma^{(\hat{\chi}_a\hat{\chi}_b)}_k+\cR_k\right)^{-1} \Gamma^{(\hat{\chi}_i\hat{\chi}_b\hat{\chi}_c)}_k \left(\Gamma^{(\hat{\chi}_c \hat{\chi}_d)}_k+\cR_k\right)^{-1} \partial_t \cR_k^{(\hat{\chi}_d \hat{\chi}_a)} \right] \right|_{\hat{\chi}=0} \, , 
\end{eqnarray}
while the flow of a two-point function has the general form
\be \label{twopointproj}
\begin{split}
	\partial_t \Gamma^{(2)}_k =& \, \text{STr} \left[  \left(\Gamma^{(2)}_k+\cR_k\right)^{-1} \Gamma^{(3)}_k \left(\Gamma^{(2)}_k+\cR_k\right)^{-1} \Gamma^{(3)}_k \left(\Gamma^{(2)}_k+\cR_k\right)^{-1} \partial_t \cR_k \right] \\
	& \; \; -\frac{1}{2} \text{STr} \left[\left(\Gamma^{(2)}_k+\cR_k\right)^{-1} \Gamma^{(4)}_k \left(\Gamma^{(2)}_k+\cR_k\right)^{-1}  \partial_t \cR_k \right] \, . 
\end{split}
\ee
Thus, determining the $k$-dependence of the expansion coefficients appearing at the $n$th order in the fluctuation fields requires knowledge about the coefficients at order $n+1$ and $n+2$. Any approximation of the exact solution has to supply this information in order to close the hierarchy at any finite order $n$. Typically, this closure is provided by identifying couplings $\bar{u}_{n,j}(k)$ appearing in the interaction vertices of different order.

\subsubsection{Vertex structures in the covariant setting}
\label{sect.3.2.2}
Before discussing the intricacies related to fluctuation field computations on a foliated spacetime, it is useful to illustrate the working of the approach in the covariant setting. Explicit computations based on this setup have been reported in \cite{Christiansen:2012rx,Christiansen:2014raa,Meibohm:2015twa,Christiansen:2015rva,Denz:2016qks,Christiansen:2017bsy,Christiansen:2017cxa,Knorr:2017fus,Knorr:2017mhu,Eichhorn:2018ydy,Eichhorn:2018akn,Eichhorn:2018nda,Burger:2019upn,Bonanno:2021squ,Knorr:2021niv,Fehre:2021eob} and reviewed in \cite{Pawlowski2021}. We focus on the gravitational field $g_{\mu\nu}$ which is decomposed into background and fluctuation fields according to
\be\label{covdec}
g_{\mu\nu} = \gb_{\mu\nu} + \hat{g}_{\mu\nu} \, . 
\ee
Most of the computations utilize a flat, Euclidean background where $\gb_{\mu\nu} = \delta_{\mu\nu}$. This has the advantage that standard momentum-space techniques are available. Derivatives within the expansion coefficients can then be replaced by the momenta of the fluctuation fields. We will adopt this choice in the sequel.

Subsequently, we consider the Einstein-Hilbert action supplemented by the harmonic gauge condition
\be\label{genfunct}
\Gamma_k[\hat{g};\gb] = - \frac{1}{16 \pi G_k} \int d^dx \sqrt{g} R + \frac{1}{32 \pi G_k} \int d^dx \sqrt{\gb} \, \gb^{\mu\nu} F_\mu F_\nu \, , 
\ee
where
\be
F_\mu = \left[\Db^\alpha \delta^\beta_{\mu} - \frac{1}{2} \, \gb^{\alpha\beta} \Db_\mu  \right] \hat{g}_{\alpha\beta} \, . 
\ee
Expanding \eqref{genfunct} in a flat background, the first non-trivial term appears at second order in the fluctuation fields
\be
\Gamma^{\rm quad}_k[\hat{g};\gb] = \frac{1}{128\pi G_k} \int \frac{d^dp}{(2\pi)^d} \, p^2 \,  \left[ \delta^{\mu\alpha} \delta^{\nu\beta} + \delta^{\mu\beta} \delta^{\nu\alpha} - \delta^{\mu\nu} \delta^{\alpha\beta}\right] \hat{g}_{\mu\nu}(p) \hat{g}_{\alpha\beta}(-p) \, . 
\ee
Comparing this result with \eqref{vertexpan} gives the expansion coefficient
\be\label{2ptseh}
\Gamma_k^{(\hat{g}\hat{g})} = \frac{1}{64\pi G_k} \, p^2 \, \left[ \delta^{\mu\alpha} \delta^{\nu\beta} + \delta^{\mu\beta} \delta^{\nu\alpha} - \delta^{\mu\nu} \delta^{\alpha\beta}\right] \, .  
\ee
The tensor structure appearing in brackets can be understood as a linear combination of the two orthogonal tensor structures
\be
\cT^{(\hat{g}\hat{g})}_1 = \frac{1}{2} \left( \delta^\mu_\alpha \delta^\nu_\beta + \delta^\nu_\alpha \delta^\mu_\beta \right) - \frac{1}{d} \delta^{\mu\nu}\delta_{\alpha\beta} \, , \qquad 
\cT^{(\hat{g}\hat{g})}_2 = \frac{1}{d} \delta^{\mu\nu}\delta_{\alpha\beta} \, , 
\ee
which can be built from a flat metric without resorting to the momentum of the field. The two $\cT$'s project a symmetric, rank-2 tensor onto its traceless and trace part, respectively. Eq.\ \eqref{2ptseh} can then be recast in the form \eqref{tensorexp},
\be
\Gamma_k^{(\hat{g}\hat{g})} = \bar{u}_2(p^2;k) \left[ \cT^{(\hat{g}\hat{g})}_1 + \frac{d-2}{2} \, \cT^{(\hat{g}\hat{g})}_2\right]
\ee
where
\be\label{ubarprop}
\bar{u}_2(p^2;k) \equiv \, \frac{1}{32 \pi G_k} \, p^2 \, . 
\ee

At this stage, the following remarks are in order. Eq.\ \eqref{ubarprop} indicates that \emph{the avatar of Newton's coupling} associated with the graviton two-point function appears at order $p^2$ in the momentum of the fluctuation field. While this coupling is derived from the gauge-fixed Einstein-Hilbert action \eqref{genfunct}, the expansion in terms of tensor structures introduced in eqs.\ \eqref{vertexpan} and \eqref{tensorexp} indicates that this coupling \emph{is different from} Newton's coupling which arises at zeroth order in the fluctuation field by evaluating \eqref{genfunct} at $\hat{g} = 0$ \cite{Manrique:2009uh,Manrique:2010am}. 

Secondly, our definition of $\bar{u}_2(p^2;k)$ promotes Newton's coupling to a form factor depending on the squared momentum $p^2$ of the fluctuation fields. The form factor then captures the non-trivial momentum dependence of the graviton two-point function.\footnote{Structurally, this is very similar to the form factors appearing in the curvature expansion of the effective average action \cite{Bosma:2019aiu,Knorr:2019atm,Draper:2020bop,Draper:2020knh} reviewed in \cite{Knorr:2021iwv,Knorr:2022dsx}.} The Wetterich equation allows to compute the dependence of $\bar{u}_2$ on $p^2$. In the covariant approach this ``reconstruction of the graviton propagator'' has been carried out in \cite{Bonanno:2021squ}. In \cite{Fehre:2021eob} it was reported that the two-point function interpolates between $\bar{u}_2 \propto \text{const}$ and $\bar{u}_2 \propto k^{\eta_h}$, $\eta_h = 0.96$, for small and large momenta, respectively.
\subsubsection{Vertex structures on a foliated spacetime}
\label{sect.3.2.3}
In comparison to the covariant setting, the foliation present in the ADM-formalism provides an additional structure which allows to distinguish the spatial and time-components of a covariant object. In particular, the squared momentum can be written as
\be\label{decp2}
p^2 = p_0^2 + \vec{p}^{\,2} \, .
\ee
It is now instructive to insert this relation into \eqref{ubarprop}
\be\label{ubarprop2}
\bar{u}_2(p_0^2, \vec{p}^{\,2};k) = \, \frac{1}{32 \pi G_k} \, \left(  p_0^2 + \vec{p}^{\,2} \right) \, . 
\ee
Generically, the coefficients in front of the two terms on the right-hand side can be different. This implies that the presence of the foliation allows to avatars of Newton's coupling associated with the spatial and ``time''-like contributions to the squared momentum
\be\label{projectionrule}
\bar{u}_2 = \frac{1}{32 \pi G_k} p^2 \qquad 
\begin{array}{ll} 
\nearrow \qquad &  \bar{u}_{2a} = \frac{1}{32 \pi G_k} \,  p_0^2 \, , \\[1.2ex] 
\searrow & \bar{u}_{2b} = \frac{1}{32 \pi G_k} \, \vec{p}^{\,2} \, . 
\end{array}  
\ee 
At the level of the action \eqref{actionansatz}, these avatars are implemented through the couplings $\alpha_1$ and $\alpha_2$. The symmetries of the covariant setting fix $\alpha_1 = \alpha_2 = 1$, indicating that the avatars generated in the split \eqref{projectionrule} should be identified.

\begin{figure}[t!]
	\centering
	\includegraphics[width=0.7\textwidth]{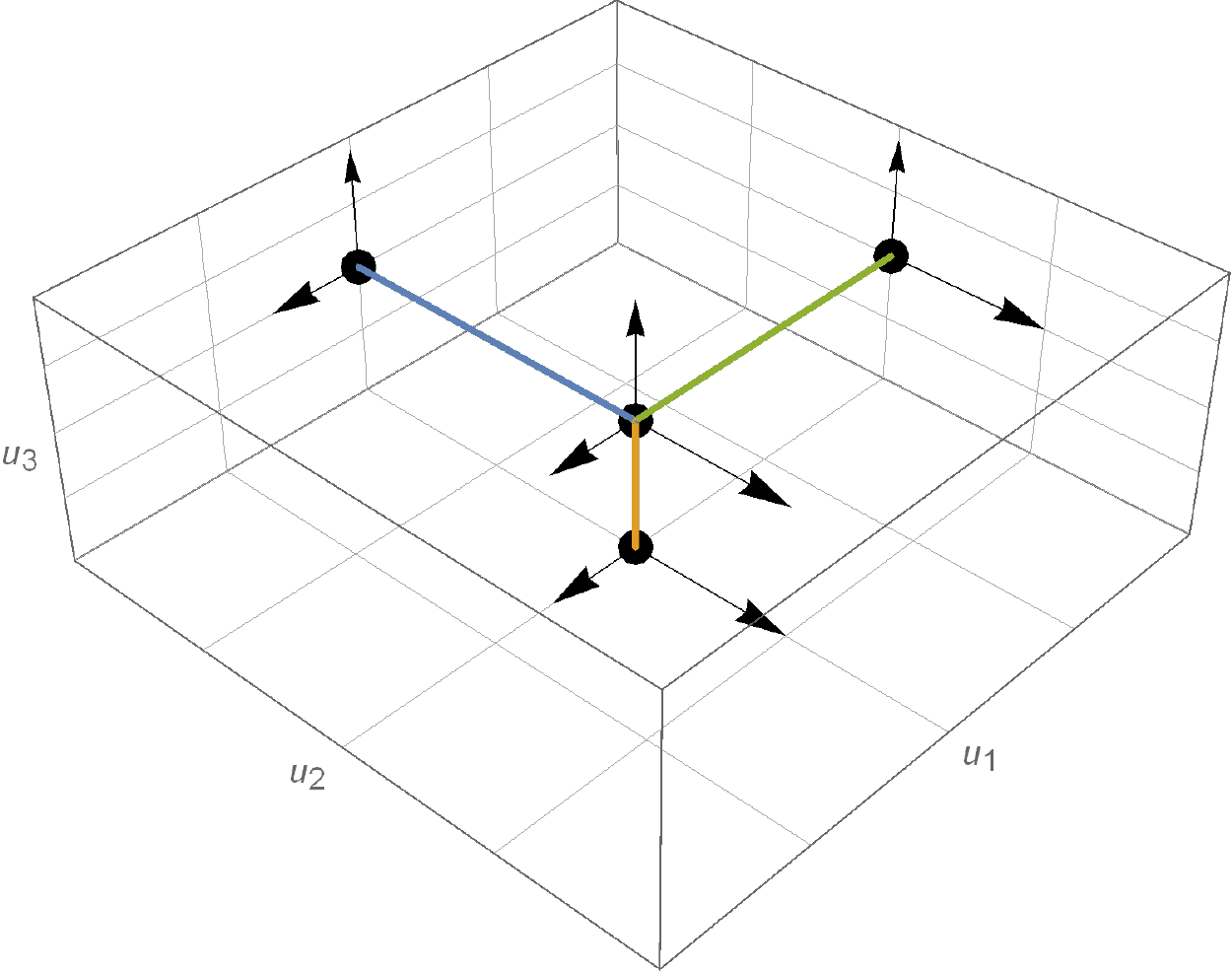}
\caption{\label{fig.1} Projection of a three-dimensional approximation of theory space to two-dimensional subspaces. The NGFP and its projections are marked by black dots. The arrows indicate eigendirections of the stability matrix, describing the RG flow in the vicinity of the fixed point. The value of the critical exponent is encoded in the length of the arrow. Notably, projections to different subspaces will see different projections of these arrows, illustrating that the critical exponents do not lend themselves to a meaningful comparison when considering subspaces spanned by different couplings a priori.}
\end{figure}

The Wetterich equation adapted to the ADM-formalism incorporates contributions which break Lorentz covariance explicitly. Hence the flow of the avatars generated in the split \eqref{projectionrule} is expected to be different. Conceptually, reading off the flow of a coupling from the terms containing $p_0^2$ or $\vec{p}^{\,2}$, should be considered as computing the beta functions of \emph{two different couplings}. The conceptual consequences of this setting are illustrated in Fig.\ \ref{fig.1}.
The figure depicts a generic RG fixed point and its projection to lower-dimensional subspaces spanned by a set of dimensionless couplings $u_i$ (black dots). The fixed point should be visible in all projections (straight lines). The stability coefficients accessed in the  projections capture different properties of the RG flow in the vicinity of the fixed point though. Thus there is no a priori reason that their values found in different projections actually agree. This applies specifically for the two projections described in eq.\ \eqref{projectionrule}. Convergence of stability coefficients requires studying extended projections which include the same subsystem. Thus, it is meaningful to compare the stability coefficients obtained in the three-dimensional and each of the two-dimensional subsystems, while a comparison among the different two-dimensional systems may not show identical stability properties.

\section{RG flows at second order in the derivative expansion}
\label{Sec.3.1}
 Upon introducing the general framework underlying fluctuation field computations in the ADM-formalism, we work out a specific example and compute the scale-dependence of the graviton two-point function resulting from the two-derivative action \eqref{actionansatz}. We start by giving the gauge-fixed propagators for the fluctuation fields in Sect.\ \ref{Sec.3.1n}. The projection of the flow on these tensor structures  and the resulting beta functions encoding the $k$-dependence of the couplings are given in Sect.\  \ref{Sec.3.2n}. Many technical details are relegated to App.\ \ref{App.A}. Throughout the computation we work in a four-dimensional flat Euclidean background and we manifestly make use of momentum-space methods in order to facilitate the computation.
 \subsection{Gauge-fixed two-point functions}
 \label{Sec.3.1n}
 We start from the action \eqref{actionansatz} which contains all terms constructed from at most two derivatives. We then adopt this action as the gravitational part of the effective average action,
 \begin{equation}\label{Gammagrav}
 	\Gamma_k^{\text{grav}}[N,N_i,\sigma_{ij}] = \frac{1}{16 \pi G_k} \int d\tau d^{3} y \, N \sqrt{\sigma} \, (\alpha_1  K^{ij}K_{ij}-\alpha_2 K^2-\leftidx{^{(d)}} R + 2\Lambda_k) \, .
 \end{equation}
 Here the couplings $\alpha_1, \alpha_2, G_k$, and $\Lambda_k$ have been promoted to depend on the coarse-graining scale $k$.  Background computations along the lines of \cite{Rechenberger:2012dt,Contillo:2013fua,Biemans:2017zca} start from a similar ansatz and subsequently evaluate the Wetterich equation \eqref{WetterichEq} at zeroth order in the fluctuation field.
 
 In our fluctuation computation, eq.\ \eqref{Gammagrav} serves as the generating functional for the tensor structures onto which we project the flow equation. In order to obtain these structures, we first supplement \eqref{Gammagrav} by the gauge-fixing and ghost terms given in eqs.\ \eqref{Gammagf} and \eqref{ghostgen} (also see App.\ \ref{App.D} for explicit expressions). We then substitute \eqref{linsplit} and expand the resulting functional in powers of the fluctuation fields in the flat background. In order to bring the Hessian $\Gamma_k^{(2)}$ into (almost) diagonal form, we implement a transverse-traceless decomposition of the fluctuation fields \cite{York1973}
 \be\label{compfields}
 \begin{split}
 	\hat{\sigma}_{ij} = & 
 	h_{ij}+ \p_i v_j + \p_j v_i + \p_i \p_j E -\frac{1}{3} \delta_{ij} \p^2 E + \frac{1}{3} \delta_{ij} \Psi,\\
 	\hat{N}_i = & u_i+ \p_i B \, , 
 \end{split}
 \ee
 where the component fields are subject to the constraints
 \be\label{compconstraints}
 \p^ih_{ij} = 0 \, , \quad \delta^{ij} h_{ij} = 0 \, , \quad \p^i v_i = 0 \, , \quad \p^i u_i = 0 \, , \quad \Psi = \delta^{ij} \hat{\sigma}_{ij} \, . 
 \ee
  These constraints ensure that the two-point functions depend on contracted spatial derivatives $\p^2 \equiv \p_i \p^i$ only \cite{Lauscher:2001ya,Benedetti:2010nr}. The field redefinition \eqref{compfields} has a non-trivial Jacobian. This extra term is conveniently accounted for by rescaling the fluctuations according to
  \begin{equation}
  	\begin{split}
  		v_i \rightarrow & \frac{1}{\sqrt{-\p^2} }v_i \, , \quad
  		E \rightarrow \frac{1}{(-\p^2)} E \, , \quad
  		B \rightarrow \frac{1}{\sqrt{-\p^2} } B \, .
  	\end{split}
  \end{equation}

The projection of the fluctuation fields $\hat{\sigma}_{ij}, \hat{N}_i$ onto the subspaces spanned by the component fields is readily achieved through projection operators. Introducing the unit on the space of symmetric $3\times 3$-matrices
\be
\mathbb{1}^{ij}{}_{kl} \equiv \frac{1}{2} \left( \delta^i_k \delta^j_l + \delta^j_k \delta^i_l \right) \, , 
\ee
the subspaces for the component fields in \eqref{compfields} are spanned by
\begin{equation}\label{projectors1}
	\begin{split}
		{\Pi_{\Psi}}^{ij}{}_{kl} &= \frac{1}{3} \sigma^{ij} \sigma_{kl} \, ,\\
		{\Pi_E}^{ij}{}_{kl} &=(\p^i \p^j -\frac{1}{3}\delta^{ij} \p^2 )(\frac{2}{3} \p^4)^{-1} (\p_k \p_l -\frac{1}{3}\delta_{kl} \p^2) \, ,\\
		{\Pi_v}^{ij}_{~~kl} &=2 \Big( \delta^{(j}_{(l} \, \p^{i)} \, \p^{-2} \, \p_{k)}
		\Big)-2 \p^i \p^j \p^{-4} \p_k \p_l,\\
		{\Pi_h}^{ij}_{~~kl} &= \mathbb{1}^{ij}_{~~kl}-{\Pi_\Psi}^{ij}_{~~~kl}-{\Pi_E}^{ij}_{~~kl}-{\Pi_v}^{ij}_{~~kl}\, .
	\end{split}
\end{equation}
Similarly, we also give the projection tensors for the space of vector fields
\begin{equation}\label{projectors2}
	\begin{split}
		{\Pi_B}{}^i{}_j  \equiv \p^i \, \p^{-2} \, \p_j \, , \qquad 
		{\Pi_u}{}^i{}_j \equiv \delta^i_j - \p^i \, \p^{-2} \, \p_j \, .
	\end{split}
\end{equation}

\begin{table}[t!]
	\renewcommand{\arraystretch}{1.5}
	\begin{center}
		\begin{tabular}{ll}
			\hline \hline
			fields $(i,j)$  & $\Gamma^{{\rm grav} \, (ij)}_{k}+\frac{1}{2}\Gamma^{{\text{gf} \, (ij)}}_{k}$ \\ \hline \hline
			$h_{ij} h^{kl}$ & $\frac{1}{32 \pi G_k} \left(  (\alpha_1 p_0^2 + \vec{p}^{\,2})-2 \Lambda_k \right) \, {\Pi_h}^{ij}_{~~kl} $ \\ \hline 
			$v_i v^j$ & 
			$\frac{1}{16 \pi G_k} \left( (\alpha_1 p_0^2 + \vec{p}^{\,2})- 2 \Lambda_k \right) {\Pi_u}{}^i{}_j$ \\
			$EE$ & 
			$\frac{1}{48 \pi G_k} \left(  (\alpha_1 p_0^2 + \vec{p}^{\,2}) - 2 \Lambda_k \right)$ \\
			$\Psi\Psi$ & 
			$- \frac{1}{192 \pi G_k} \left( ( 6 \alpha_2 -2 \alpha_1 - 3) p_0^2 +\vec{p}^{\,2} -  2 \Lambda_k \right)$ \\
			$\hat{N}\hat{N}$ & 
			$\frac{1}{16 \pi G_k} (p_0^2+\vec{p}^{\,2})$ 	\\ 
			$\Psi \hat{N}$ & 
			$-\frac{1}{16 \pi G_k} \left(p_0^2+ \vec{p}^{\,2} -2 \Lambda_k \right)$ \\
			$u^i u_j$ & 
			$\frac{1}{16 \pi G_k} (p_0^2+\alpha_1 \vec{p}^{\,2}) {\Pi_u}{}^i{}_j $ 	\\ 
			$BB$ & 
			$\frac{1}{16 \pi G_k} \left( p_0^2+ (1 + 2 \alpha_1 - 2 \alpha_2) \, \vec{p}^{\,2} \right)$ \\ \hline
			$\bar{c}c$ & $\sqrt{2} \, (p^2_0 +\vec{p}^{\,2})$ \\
			$\bar{b}^i b_i$ & $\sqrt{2} \, (p^2_0 +\vec{p}^{\,2}) \, \Pi_u{}_j{}^i$
			\\ \hline \hline
		\end{tabular}
	\end{center}
	\caption{\label{table.hessian} Matrix elements of the Hessian $\Gamma^{(2)}_k$. The first line is singled out since this is the tensor structure onto which we are going to project the flow equation. The second block gives the propagators associated with the metric degrees of freedom while the third block captures the information about the propagators in the ghost sector.}
\end{table}
At this point we have all the ingredients to write down the matrix elements of the Hessian $\Gamma_k^{(2)}$. Including the contribution of the gauge-fixing terms, these are tabulated in Table \ref{table.hessian}. At this point a closer inspection of these expressions in order. All two-point functions are proportional to the projection operators restricting the fluctuation fields to the corresponding transverse (and traceless) subspaces. For scalars, this projection is trivial and the corresponding projectors are omitted. For $\alpha_1 = \alpha_2 = 1$, the gauge-fixed (inverse) propagators all come with a \emph{relativistic dispersion relation}. In this case, the spatial and time-components of the momentum combine into a relativistic four-momentum $p_0^2 + \vec{p}^{\, 2} = p^2$. This singles out the gauge-fixing adopted in eq.\ \eqref{gaugechoice} \cite{Biemans:2016rvp}. The cosmological constant then plays the role of a mass term in the two-point function.

At this point we have all the ingredients to specify the components of the regulator $\cR_k(\Box)$. Following the nomenclature of \cite{Codello:2008vh}, we implement a Type I regularization, which fixes $\cR_k$ through the substitution rule
\begin{equation}\label{eq.regdef}
	\Box \rightarrow P_k (\Box) \equiv \Box + R_k(\Box) \, ,
\end{equation}
where $R_k(\Box)$ is a scalar cutoff function. Throughout this work, we chose the cutoff function to be of Litim-type \cite{Litim:2000ci,Litim:2001up}, 
\be\label{eq.litim}
R_k (\Box) =(k^2-\Box)\Theta (k^2-\Box) \, ,
\ee
where $\Theta(x)$ is the Heaviside step function. Inspecting Table \ref{table.hessian}, one identifies various candidates for the coarse-graining operator $\Box$. Adopting a momentum-space representation, these are within the class
\be\label{eq.coarsegrainingop}
\Box_\alpha = \alpha \, p_0^2 + \vec{p}^{\,2} \, ,
\ee
where $\alpha$ depends on $k$. Based on the kinetic terms for the various component fields, we would then encounter as many different forms of $\Box_\alpha$ as there are distinct   dispersion relations. 

 The $\alpha$-dependence of $\Box$ introduces a significant number of technical complications in the computation. This can be understood from noticing that each operator $\Box_\alpha$ leads to a different stepfunction $\Theta (k^2-\Box_\alpha)$. Loop integrals then involve different step functions which results in quite complicated domains when integrating over loop momenta \cite{Becker:2017tcx}. We avoid this complication by employing a modified (but perfectly admissible) replacement rule for the regularization
 \be\label{eq.TypeIJian}
 \Box_\alpha \rightarrow \Box_\alpha + R_k(\Box_{\alpha = 1}) \, . 
 \ee
This results in the matrix elements for $\cR_k$
\be\label{eq.TypeIJianexamples}
\begin{split}
\cR_k^{hh} = & \frac{1}{32 \pi G_k} R_k(p^2) \, \Pi_h{}^{ij}_{kl} \, , \qquad
\cR_k^{uu} =  \frac{\alpha_1}{32 \pi G_k} R_k(p^2) \, \Pi_u{}^{i}_{j} \, ,
\end{split}
\ee
and similar for the other matrix entries. We will adopt the choice \eqref{eq.TypeIJian} in the sequel.\footnote{In principle, one can chose other regularization procedures as well. These differ by the anomalous dimension of the coupling appearing in $\p_k \cR_k$. For instance, one could adopt
	\begin{equation} \label{eq.TypeImod}
		\Box_\alpha \rightarrow P_k (\Box_{\alpha =1}) + \left(\Box_\alpha - \Box_{\alpha = 1} \right) \, .
	\end{equation}
	In comparison to \eqref{eq.TypeIJian} the two choices differ by terms proportional to the beta functions associated with $\alpha_1$ and $\alpha_2$. These extra contribution vanish on a fixed point by definition, so that eq.\ \eqref{eq.TypeImod} and \eqref{eq.TypeIJian} lead to the same fixed point structure.
}

The different dispersion relations listed in Table \ref{table.hessian} then lead to the same profile function $R_k$. This function is just a function of the four-momentum $p^2$ and does not include $k$-dependent couplings. In this way, the argument of $R_k$ is covariant and preserves Lorentz symmetry. In this way, the choice of regulator minimizes the Lorentz-symmetry violating terms generated by the regularization procedure. Still the non-linearity of the field decomposition \eqref{linsplit} makes the regularization procedure non-covariant.

\subsection{Projecting and closing the flow equation}
\label{Sec.3.2n}
At this point, we have all the ingredients for specifying the projection of the Wetterich equation underlying our computation. Specifically, we project the flow onto the graviton two-point function singled out in the first line of Table \ref{table.hessian}. Substituting this expression into the left-hand side of the flow equation gives
\be\label{projectionlhs}
32 \pi \, \p_t \Gamma^{(hh)}_k(p_0^2,\vec{p}^{\,2}) = 
 \partial_t \left( \frac{\alpha_1}{G_k} \right)  \, p_0^2 \, \Pi_h + \partial_t \left( \frac{1}{G_k} \right) \, \vec{p}^{\,2} \, \Pi_h -2 \partial_t \left( \frac{\Lambda_k}{G_k} \right) \Pi_h \, .
\ee
This indicates that the projection can track up to three scale-dependent couplings $\alpha_1$, $\Lambda_k$, and $G_k$.\footnote{In principle, the formalism can also be used to track the scale-dependence of $\alpha_2$. This requires considering one additional two-point function, e.g., $\Gamma^{(\Psi\Psi)}_k(p_0^2,\vec{p}^{\,2})$ which contains the corresponding coupling. While this computation is conceptually straightforward, it essentially doubles the complexity of the underlying algebra. For this reason, we leave this extension for future work.} This projection then identifies the tensor structures which need to be extracted from the trace on the right-hand side of the equation. We define
\be\label{projectionrhs}
\begin{split}
\frac{1}{2} {\rm STr}\left[\cdots\right] \simeq 
	& \, T_{p_0} (G_k, \alpha_1, \alpha_2,\Lambda_k, \eta_N,\partial_t \alpha_1) \, p_0^2 \, \Pi_h  \\ 
	& \, + T_{\vec{p}} (G_k, \alpha_1, \alpha_2,\Lambda_k,\eta_N, \partial_t \alpha_1) \,  \vec{p}^{\,2} \, \Pi_h \\
& \, -T_{0}  (G_k, \alpha_1, \alpha_2,\Lambda_k,\eta_N, \partial_t \alpha_1) \, \Pi_h ,
\end{split}
\ee
where the $\simeq$ indicates that we neglect terms outside of the projection subspace and the sign in front of $T_0$ has been adjusted for later convenience. The arguments of the traces highlight the dependence of the contributions on the couplings and we defined the anomalous dimension of $G_k$ as $\eta_N = \partial_t G_k / G_k$. Equating the coefficients multiplying the independent tensor structures in eqs.\ \eqref{projectionlhs} and \eqref{projectionrhs} leads to the following system of first order differential equations
\be
\begin{split}
\frac{1}{32 \pi} \partial_t \left( \frac{\alpha_1}{G_k} \right) & \, = T_{p_0} (G_k, \alpha_1, \alpha_2,\Lambda_k,\eta_N,\partial_t \alpha_1) \, ,\\
\frac{1}{32 \pi} \partial_t \left( \frac{1}{G_k} \right) & \, = T_{\vec{p}} (G_k, \alpha_1, \alpha_2,\Lambda_k,\eta_N,\partial_t \alpha_1) \, ,\\
\frac{1}{16 \pi} \partial_t \left( \frac{\Lambda_k}{G_k} \right) & \, = T_{0} (G_k, \alpha_1, \alpha_2,\Lambda_k,\eta_N,\partial_t \alpha_1) \, .
\end{split}
\ee
As indicated above, the projection does not capture information about the running of $\alpha_2$. So this coupling has to be treated as a dimensionless parameter.

At this point the computation has reduced to determining the coefficients $T$. Inspecting eq.\ \eqref{twopointproj} shows that this requires information about the ($k$-dependent) $3$- and $4$-point vertices of the theory. Generically, this information is not available exactly, and one has to adopt an approximation in order to close the system of equations. We then  generate the relevant $3$- and $4$-point vertices by taking additional functional derivatives of $\Gamma_k^{\rm grav} + \Gamma_k^{\rm ghost}$ with respect to the collection of fluctuation fields $\hat{\chi}$ and subsequently specifying to the flat background. Note that the projection to \eqref{projectionlhs} entails that is is not necessary to consider all vertices: only $3$-point vertices with at least on leg being $h_{ij}$ and $4$-point vertices with two legs associated with the field $h_{ij}$ contribute. Our actual computation of the coefficients $T$ retains the full momentum dependence of these vertices on the loop-momentum. While the actual computation is conceptually straightforward, it is technically involved and we collect the details in App.\ \ref{App.B}.

The $k$-dependence of the couplings is then conveniently encoded in the beta functions for the dimensionless couplings $(g_k,\lambda_k,\alpha_1)$
\be\label{betafctsdef}
\partial_t g_k = \beta_{g_k}(g_k,\lambda_k,\alpha_1;\alpha_2) \, ,  \quad \partial_t \lambda_k = \beta_{\lambda_k}(g_k,\lambda_k,\alpha_1;\alpha_2) \, ,  \quad \partial_t \alpha_1 = \beta_{\alpha_1}(g_k,\lambda_k,\alpha_1;\alpha_2) \, ,  
\ee
 with
\be\label{dimlessdef}
g_k \equiv G_k \, k^2 \, , \quad \lambda_k \equiv \Lambda_k \, k^{-2} \, .  
\ee
The parametric dependence on $\alpha_2$ is highlighted by the semicolon.

The explicit expressions for the beta functions \eqref{betafctsdef} are rather lengthy and can be provided in form of an auxiliary {\tt Mathematica} notebook upon request. Here we limit ourselves to giving the results for the foliated Einstein-Hilbert action ($\alpha_1=\alpha_2=1$) only. In this case we find
\be\label{betalambdaEH}
\begin{split}
	\beta_g = & \, \left(2 + \eta_N \right) g \, , \\
\beta_\lambda = & \, (\eta_N - 2) \lambda + \frac{g}{\pi}\left\{ \frac{p^1_\lambda(\lambda) + \eta_N \, \tilde{p}^1_\lambda(\lambda)}{24 \left(1-2 \lambda \right)^2 \left(2-3 \lambda\right)^2} + \frac{p^2_\lambda(\lambda) + \eta_N \, \tilde{p}^2_\lambda(\lambda)}{7200 \left(1-2 \lambda \right)^3 \left(2-3 \lambda\right)^3} \right\} \, .
\end{split}
\ee
The polynomials are tabulated in the first block of Table \ref{Tab.2}.
\begin{table}[t!]
	\renewcommand{\arraystretch}{1.5}
\begin{center}
\begin{tabular}{ll}
\hline\hline
$p^1_\lambda(\lambda)$ & $12 \left(18 \lambda^4-116 \lambda^3+142 \lambda^2-61 \lambda+8\right)$\\
$\tilde{p}^1_\lambda(\lambda)$ & $-27 \lambda^4+211 \lambda^3-259 \lambda^2+106 \lambda-12$\\
$p^2_\lambda(\lambda)$ & $10 \left(68040 \lambda ^6+439308 \lambda ^5-1156914 \lambda ^4+1012717 \lambda ^3-419962 \lambda ^2+94584 \lambda -11760\right)$ \\
$\tilde{p}^2_\lambda(\lambda)$ & $87480 \lambda ^6-1420428 \lambda ^5+2828930 \lambda ^4-2242429 \lambda ^3+851098 \lambda ^2-166200 \lambda +17520$ \\ \hline
\multirow{2}*{$p^3_\lambda(\lambda)$} & $408240 \lambda ^8-2262816 \lambda ^7+6128784 \lambda ^6-10347048 \lambda ^5$ \\
&  $+10788945 \lambda ^4-6629544 \lambda ^3+2229588 \lambda ^2-344176 \lambda +12704$ \\
$\tilde{p}^3_\lambda(\lambda)$ & $4968 \lambda ^4-67980 \lambda ^3+94335 \lambda ^2-40380 \lambda +4148$ \\ \hline
$p^4_\lambda(\lambda)$ & $88248 \lambda ^5-180807 \lambda ^4+125287 \lambda ^3+3744 \lambda ^2-30324 \lambda +6376$ \\
$\tilde{p}^4_\lambda(\lambda)$ & $20055 \lambda ^4-32850 \lambda ^3+19676 \lambda ^2-771 \lambda -2158$  
\\\hline \hline
\end{tabular}
\caption{\label{Tab.2} Polynomials appearing in the beta functions \eqref{betalambdaEH} and the anomalous dimension \eqref{etaN}.}
\end{center}
\end{table}
The anomalous dimension takes the form
\be\label{etaN}
\eta_N = \frac{g B_1(\lambda)}{1-g B_2(\lambda)} \, . 
\ee

\begin{figure}
	\includegraphics[width=0.48\textwidth]{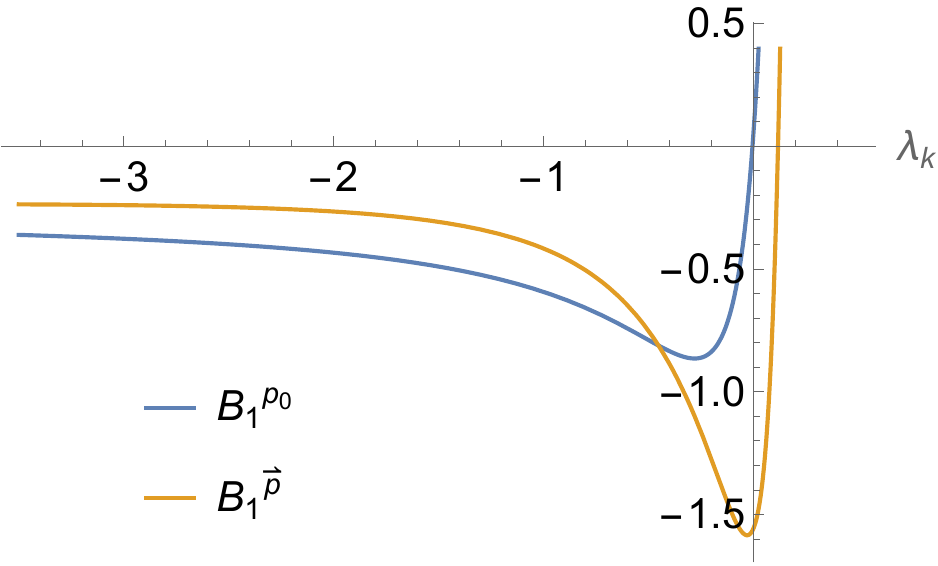}~~~~~~
	\includegraphics[width=0.48\textwidth]{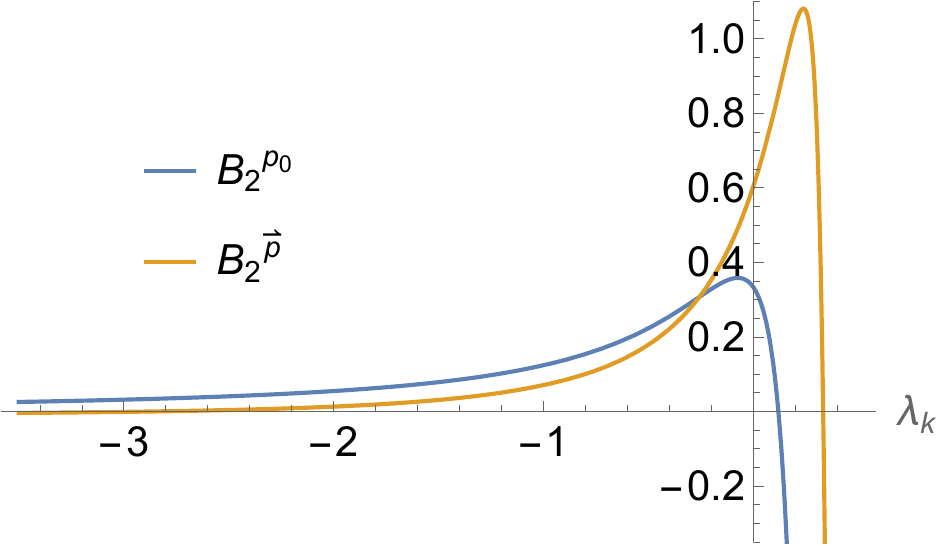}
	\caption{\label{Fig.Bcomp} Comparison between the functions $B_1(\lambda)$ (left panel) and $B_2(\lambda)$ (right panel) obtained from reading off the anomalous dimension $\eta_N$ from the time- (blue lines) and spatial component (orange lines) of the momentum.}
\end{figure}

In the foliated Einstein-Hilbert case, eq.\ \eqref{projectionlhs} indicates that the functions $B_1(\lambda)$ and $B_2(\lambda)$ can either be obtained from the coefficient of the $p_0^2$-term ($p_0$-projection) or the $\vec{p}^{\,2}$-term ($\vec{p}^{}$-projection). The $p_0$-projection yields
\be\label{Bq0proj}
\begin{split}
B_1^{p_0}(\lambda) = &\frac{48 \lambda ^2-60 \lambda +19}{2 \pi  (1-2 \lambda)^2 (2-3 \lambda )^2} -\frac{p^3_\lambda(\lambda)}{360 \pi  (1-2 \lambda)^4 (2-3 \lambda)^4} \, ,\\
B_2^{p_0}(\lambda) = & -\frac{48 \lambda ^2-60 \lambda +19}{12 \pi  (1-2 \lambda)^2 (2-3 \lambda)^2} +\frac{\tilde{p}^3_\lambda(\lambda)}{720 \pi  (1-2 \lambda)^3 (2-3 \lambda)^2}\, . \\	
\end{split}
\ee
the projection on the spatial momentum yields an extra contribution to $B_1(\lambda)$ and $B_2(\lambda)$:
\be\label{Bqvecproj}
\begin{split}
	B_1^{\vec p}(\lambda) = & B_1^{p_0}(\lambda) -\frac{p^4_\lambda(\lambda)}{315 \pi  (2-3 \lambda )^2 (1-2 \lambda )^4} \, , \\
	B_2^{\vec p}(\lambda) = & B_2^{p_0}(\lambda) - \frac{\tilde{p}^4_\lambda(\lambda)}{630 \pi  (2-3 \lambda )^2 (1-2 \lambda)^3} \, .  \\	
\end{split}
\ee
 The functions $B_1$ and $B_2$ obtained from the two projections are compared in Fig.\ \ref{Fig.Bcomp}. While the results differ quantitatively, the functions agree on a qualitative level. Together with the generalization of our results including the beta function for $\alpha_1$ and the dimensionless parameter $\alpha_2$ provided in the supplementary material, eqs.\ \eqref{betalambdaEH}-\eqref{Bqvecproj} constitute the main result of this section.

We stress that the anomalous dimension $\eta_N$, eq.\ \eqref{etaN}, is conceptually different from the ``dynamical anomalous dimension'' $\eta_h$ considered in the covariant approach. Determining the latter requires information about the scale-dependence of the graviton $3$-point vertex in order to disentangle the running of the graviton self-interaction from the wave function renormalization of the fluctuation field. This information is not contained in the present projection though. As a consequence, eq.\ \eqref{betalambdaEH} dictates that $\eta_N(g_*,\lambda_*) = -2$ at a NGFP.

We close the section with a conceptual remark. The differences appearing in the $p_0$- and $\vec{p}$-projection can be traced back to the $3$- and $4$-point vertices which depend on $p_0$ and $\vec{p}$ in a non-relativistic way. A prototypical example is the $4$-point vertex $\Gamma^{(hh\hat{N}\hat{N})}_k$.  Since the potential terms in the action \eqref{Gammagrav} are linear in $N$ these do not contribute to this vertex. Considering the kinetic part of the action, the definition of the extrinsic curvature entails that this sector gives non-trivial contributions which contain just the time-derivatives of the spatial metric.  Schematically, the vertex then comes with the momentum structure
\begin{equation}\label{4ptexample}
	\Gamma^{(hh\hat{N}\hat{N})}_{k} h^2 \hat{N}^2 \propto \, p_0 ^2 \, h^{kl}(-p) h_{kl}(p)\hat{ N}(-q) \hat{ N}(q)  ,
\end{equation}
where $p=(p_0,\vec{p})$ is the external 4-momentum, and $q=(q_0,\vec{q})$is the momentum running in the loop. So the vertex \eqref{4ptexample} contributes to the $p_0$-projection while it does not enter into the $\vec{p}$-projection.

This illustrates that, in general, the $n$-point functions arising in the ADM-framework are non-relativistic. At the level of the $2$-point function this difference has been eliminated by making the specific choice for the gauge-fixing term \eqref{gaugechoice}.  The Lorentz invariance breaking effect comes from the non-linear ADM-decomposition of the metric. It is then tempting to try to dispel these effects by admitting fluctuation terms of higher order in the gauge-fixing procedure. An investigation along these lines will be left for the future work. 
 
\section{Properties of the renormalization group flow}
\label{RGanaly}
 We proceed by analyzing the fixed point structure and phase diagrams entailed by the beta functions \eqref{betafctsdef}. We start with the foliated Einstein-Hilbert truncation, where $\alpha_1=\alpha_2=1$ in Sect.\ \ref{Sec.5.1}. The flow of the full system including the beta function for $\alpha_1$ is discussed in Sect.\ \ref{Sec.5.2}. As the main result, we show that all settings exhibit a non-Gaussian fixed point (NGFP) suitable for providing the high-energy completion of the RG flow. 
 \subsection{The foliated Einstein-Hilbert truncation}
 \label{Sec.5.1}
 The beta functions for the foliated Einstein-Hilbert truncation are given in eq.\ \eqref{betalambdaEH}. Depending on whether the anomalous dimension of Newton's coupling $\eta_N$ is read off from the  time or spatial component of the momentum appearing in the graviton two-point function, the functions $B_1(\lambda)$ and $B_2(\lambda)$ are given in eqs.\ \eqref{Bq0proj} and \eqref{Bqvecproj}, respectively. Following the terminology introduced in the last section, we refer to these cases as the $p_0$-projection and the $\vec{p}$-projection, respectively.
 
 \begin{table}
 	\begin{center}
 		\begin{tabular}{| c | c | c | c  c  | c  c |  }
 			\hline \hline
 		Work & Projection & Fixed Points  &\multicolumn{2}{c|}{Couplings} & \multicolumn{2}{c|}{Critical Exponents}\\	
 			\hline	
 			~ &~ & & $g_*$ & $\lambda_*$ &  $\theta_1$ & $\theta_2$  \\ \hline \hline
 	\multirow{4}{*}{this article}	&	\multirow{4}{*}{$p_0$-projection} & GFP & $0$ & $0$ & $2$ & $-2$ \\ \cline{3-7}
 		&	~&NGFP$_1$ & $1.43$  & $-0.12$          &  \multicolumn{2}{c |}{$4.42\pm1.38 i $}\\ \cline{3-7}
 		&	~&NGFP$_2$ & $-0.40$  & $0.14$  &  \multicolumn{2}{c |}{$1.41\pm3.84 i $}\\
 		&	~&NGFP$_3$ & $5.86$  & $2.24$          & $9.37$  & $4.52$ \\
 			\hline  \hline
 			\multirow{4}{*}{this article}	&	\multirow{4}{*}{$\vec{p}$-projection}  & GFP & $0$ & $0$ & $2$ & $-2$ \\ \cline{3-7}
 		&	~ & NGFP$_1$ & $0.78$  & $-0.07$          & $5.01$ & $2.73$\\ \cline{3-7}
 		&	~& NGFP$_2$ &  $-0.23$ &   $0.22$  &  \multicolumn{2}{c |}{$-0.80\pm5.06 i $}\\
 		&	~& NGFP$_3$ & $0.36$  & $1.04$          &$27.50$ &$-5.17$\\
 			\hline \hline
 			Ref.\ \cite{Manrique:2011jc} & ADM & NGFP & $0.19$ & $0.31$ & \multicolumn{2}{c |}{$1.07 \pm 3.31i$} \\ 
 			Ref.\ \cite{Biemans:2017zca} & background level & NGFP & $0.90$ & $0.24$ & \multicolumn{2}{c |}{$1.06 \pm 3.07i$} \\ \hline
 			\multirow{2}{*}{Ref.\ \cite{Christiansen:2014raa}} & covariant & 
 			\multirow{2}{*}{NGFP} & \multirow{2}{*}{$0.61$}  & \multirow{2}{*}{$0.32$} & \multicolumn{2}{c|}{\multirow{2}{*}{$1.27 \pm 3.01i$}} \\
 			& $3$-point function & &   && &   \\ \hline \hline
 		\end{tabular}
 	\end{center}
 	\caption{ Summary of the fixed point structure obtained from the foliated Einstein-Hilbert truncation using the $p_0$-projection (top box) and $\vec{p}$-projection (middle box). The NGFP$_1$ is present in both projection schemes and serves as a UV-attractor for the RG flow in its vicinity. The properties of the NGFP obtained within the background computations carried out within the ADM-framework \cite{Manrique:2011jc,Biemans:2017zca} and the covariant fluctuation computation of \cite{Christiansen:2014raa} are added for comparison. }\label{TableFPGLam}
 \end{table}
 \emph{Fixed points.} We first identify the fixed points $(g_*, \lambda_*)$ where, by definition, $\beta_g(g_*, \lambda_*) = 0, \beta_\lambda(g_*, \lambda_*) = 0$ and obtain their stability properties by computing the critical exponents from the stability matrix \eqref{stabmat}. We first observe that both projection schemes admit a GFP
 \be\label{EH-GFP}
 {\rm GFP:} \qquad (g_*,\lambda_*) = (0,0) \, , \qquad \theta_1 = 2 \, , \; \; \theta_2 = - 2 \, . 
 \ee
 The critical exponents agree with the ones obtained from canonical power counting, in agreement with the definition of a GFP advocated in \cite{Reuter:2019byg,Saueressig:2023irs}. The GFP is a saddle point in the $g$-$\lambda$-plane. The eigenvector associated with the UV-repulsive eigendirection indicates that this fixed point cannot serve as a UV-completion for RG-trajectories with $g_k > 0$.
 
 In addition to the GFP, our numerical search for real roots also revealed three NGFPs, present in both the $\vec{p}$- and $p_0$-projection. Their properties are summarized in Table \ref{TableFPGLam}. The fixed point of physical interest is NGFP$_1$. It is located at $g_* > 0$ and $\lambda_* < 0$ and serves as a UV-attractor for the RG flow in its vicinity. The negative sign of $\lambda_*$ indicates that the dimensionless graviton mass appearing in the two-point function at the fixed point is positive. Comparing the results for NGFP$_2$ and NGFP$_3$, it is evident that the two projections do not necessarily give rise to the same qualitative behavior. It is therefore remarkable that the NGFP$_1$ found in both cases share the same qualitative features.
 
 The last three lines summarize properties for NGFPs reported from either ADM-based computations at the background level \cite{Manrique:2011jc,Biemans:2017zca} or flows of the graviton two-point function studied in the covariant approach \cite{Christiansen:2014raa}. Strictly speaking, these results do not lend themselves to a direct comparison since - from the perspective of the fluctuation field approach - the couplings $g$ and $\lambda$ listed in the various blocks are associated with different correlation functions and projection prescriptions. Nevertheless, the results form a coherent picture in the sense that they all point towards the existence of a phenomenologically interesting NGFP in foliation approach. Qualitatively, the features of this fixed point agree with the ones observed in fluctuation computations carried out in the covariant framework.

\begin{figure}
 	\centering
 	\includegraphics[width = 0.48\textwidth]{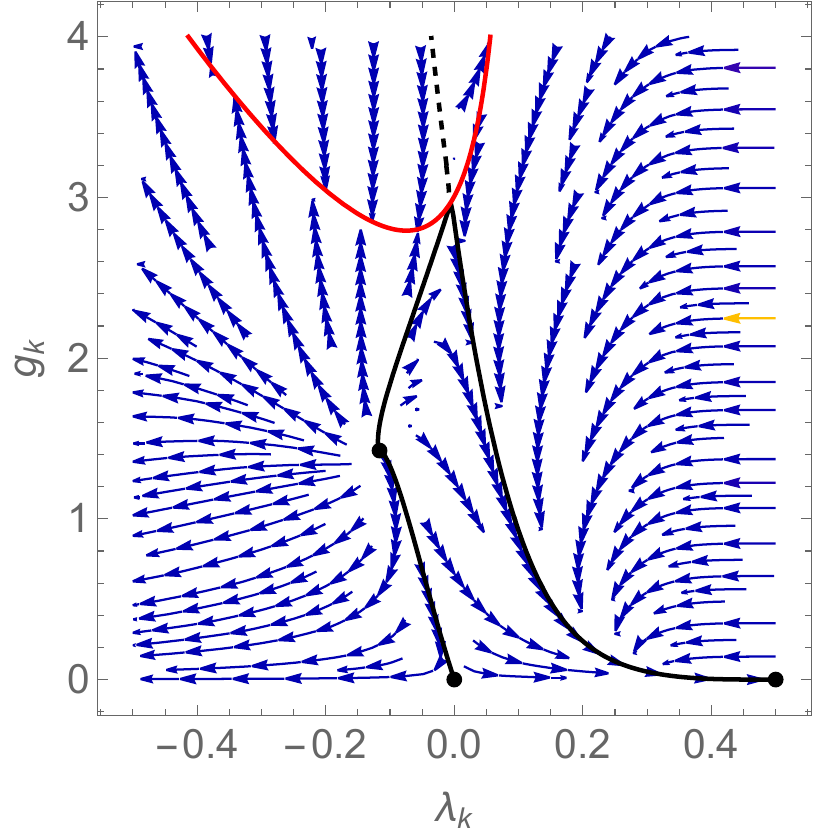} \; 
 	\includegraphics[width = 0.48\textwidth]{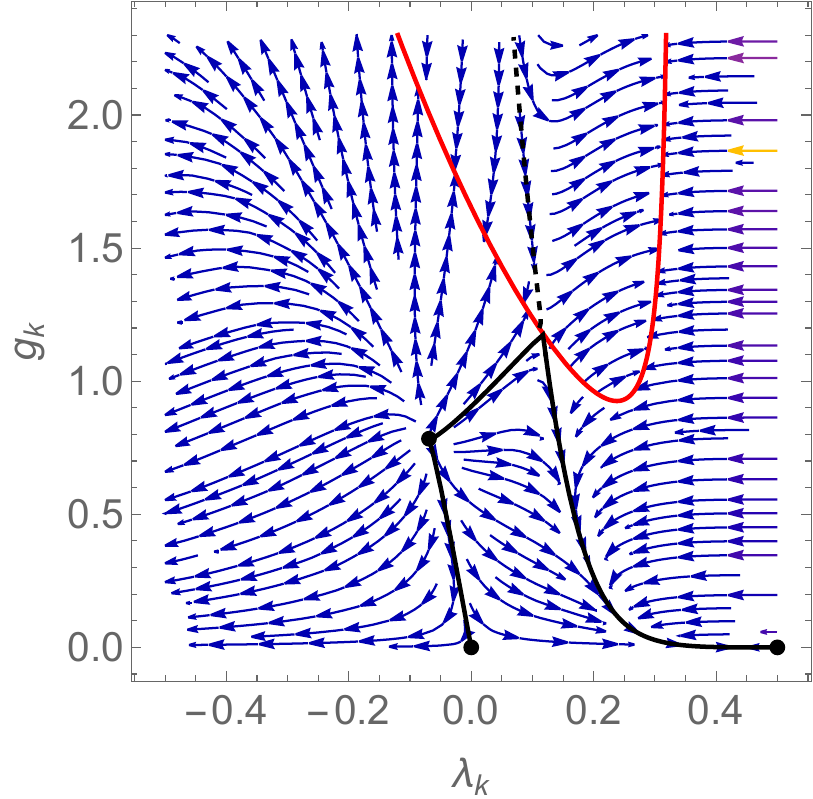}
 	\caption{\label{FlowDiaFEH} Phase diagram obtained from the beta functions of the foliated Einstein-Hilbert truncation in the $p_0$-projection (left diagram) and the $\vec{p}$-projection (right diagram) with the arrows pointing towards lower values of the coarse-graining scale $k$. The diagrams focus on the physically interesting region where $g_k \ge 0$. The GFP \eqref{EH-GFP}, the NGFP$_1$ given in Table \ref{TableFPGLam}, and the IR-FP \eqref{IR-FP} are marked by the black dots while the red lines mark a singular locus of the beta functions. The flow is governed by the interplay of these fixed points. The RG trajectory connecting the NGFP$_1$ in the UV to the GFP in the IR (Separatrix) is highlighted by the black line. The other black lines have been added to highlight the boundary of the region where the RG trajectories interpolate between NGFP$_1$ in the UV and the IR-FP as $k\rightarrow 0$.}
 \end{figure}

 \emph{Phase diagram.}  The phase diagram for the foliated Einstein-Hilbert truncation is shown in Fig.\ \ref{FlowDiaFEH}. The plots show that the $p_0$- and $\vec{p}$-projection lead to results which are qualitatively identical.
 In order to understand the structure, we first note that the line $g=0$ separates the trajectories with positive and negative $g_k$ and cannot be crossed. Moreover, there is a singular locus at $\lambda = 1/2$ linked to a singularity in the beta functions. The beta functions give rise to a second, non-trivial singular locus which is depicted by the red line. We then limit the discussion to the physically interesting region $g \ge 0$, $\lambda \le 1/2$, and the region below the red line. The RG flow in this region is governed by the interplay of the NGFP$_1$, the GFP, and the IR-fixed point
 \be\label{IR-FP}
 \text{IR-FP:} \qquad (g_*^{\rm IR}, \lambda_*^{\rm IR}) = (0, 1/2) \, . 
 \ee
 The NGFP$_2$ and NGFP$_3$ are separated from this region by the lines discussed above and do not influence the flow in this region. As indicated by the critical exponents, the NGFP$_1$ serves as a UV-attractor for the RG trajectories in its vicinity. At the IR-FP, the beta functions are ambiguous. Investigating the scaling properties of RG trajectories in its vicinity, one finds that 
 \be\label{IR-FP-scaling}
 \lambda^{\rm IR}_* - \lambda(t) \sim c_1 \, e^t \, , \qquad g(t) - g_*^{\rm IR} \sim c_2 \, e^{4t} \, ,
 \ee
 where $c_1$ and $c_2$ are constants depending on the RG trajectory chosen and we use the symbol $\sim$ to indicate that the relations are valid in the scaling regime of the IR-FP.
 
  Among the set of trajectories emanating from NGFP$_1$, two play a distinguished role in determining the structure of the phase diagram. These are highlighted by the black solid lines. Firstly, we have one trajectory which connects the NGFP in the UV to the GFP in the IR. This line has been called separatrix in \cite{Reuter:2001ag}. Secondly, there is a ``broken'' trajectory which connects the NGFP to a specific point on the singular line (red curve) and subsequently to the IR-FP in the IR. These black lines bound different phases encountered in the phase diagram. RG trajectories to the left of the separatrix (and not terminating in the red, singular line) flow to $(g,\lambda) = (0,-\infty)$ and lead to negative values $\Lambda_k$ in the IR. In this case, the mass appearing in the two-point function is positive. Trajectories in the triangle formed by the connecting lines approach the IR-FP in the IR. They all lead to a zero value of the squared graviton mass $\mu^2 \equiv - 2 \Lambda_0$. 
  
\begin{figure}
	\centering
	\includegraphics[width = 0.48\textwidth]{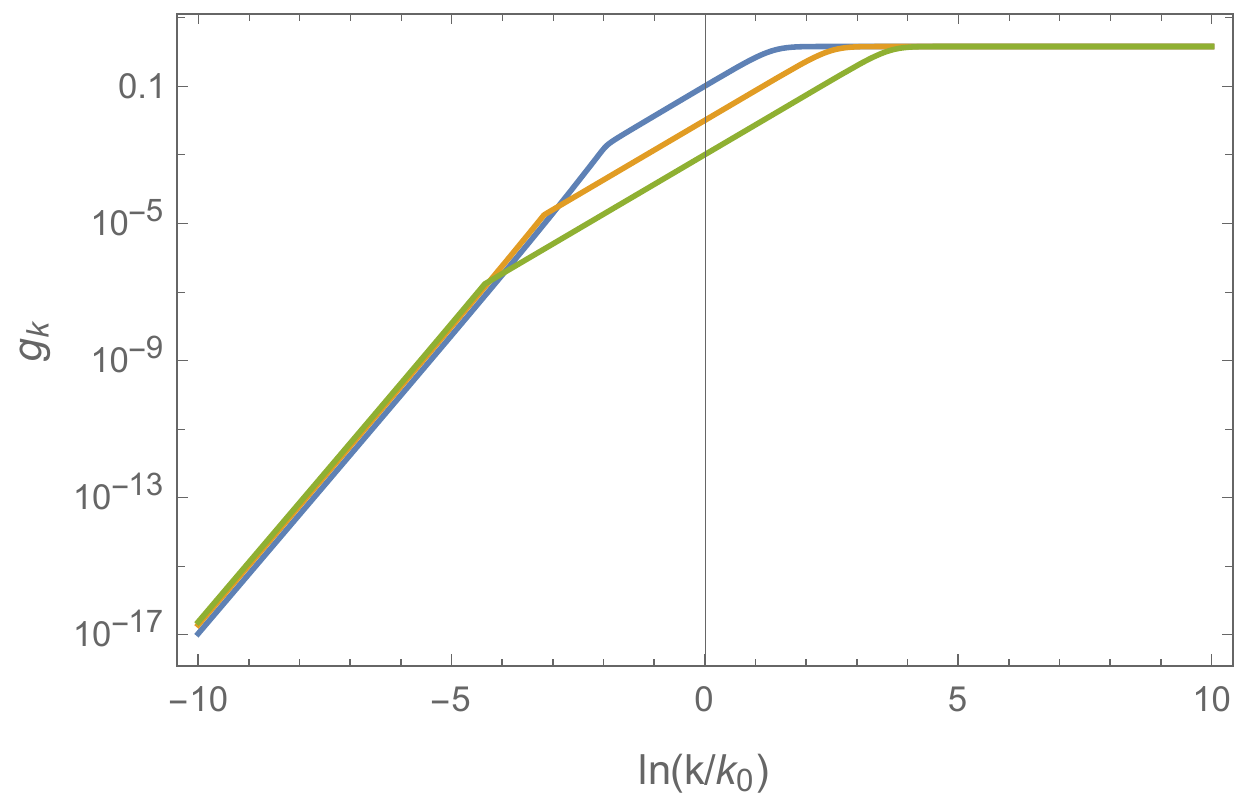} \; 
	\includegraphics[width = 0.48\textwidth]{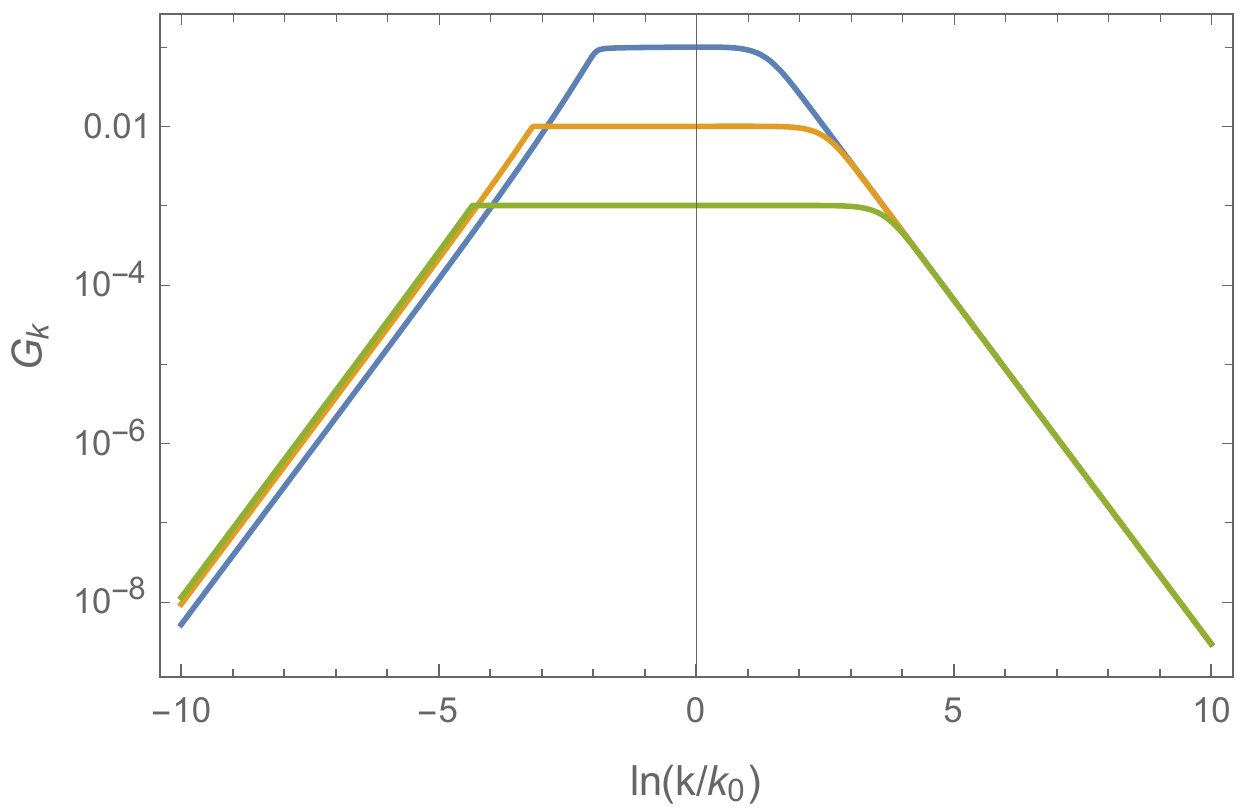} \\
	\includegraphics[width = 0.48\textwidth]{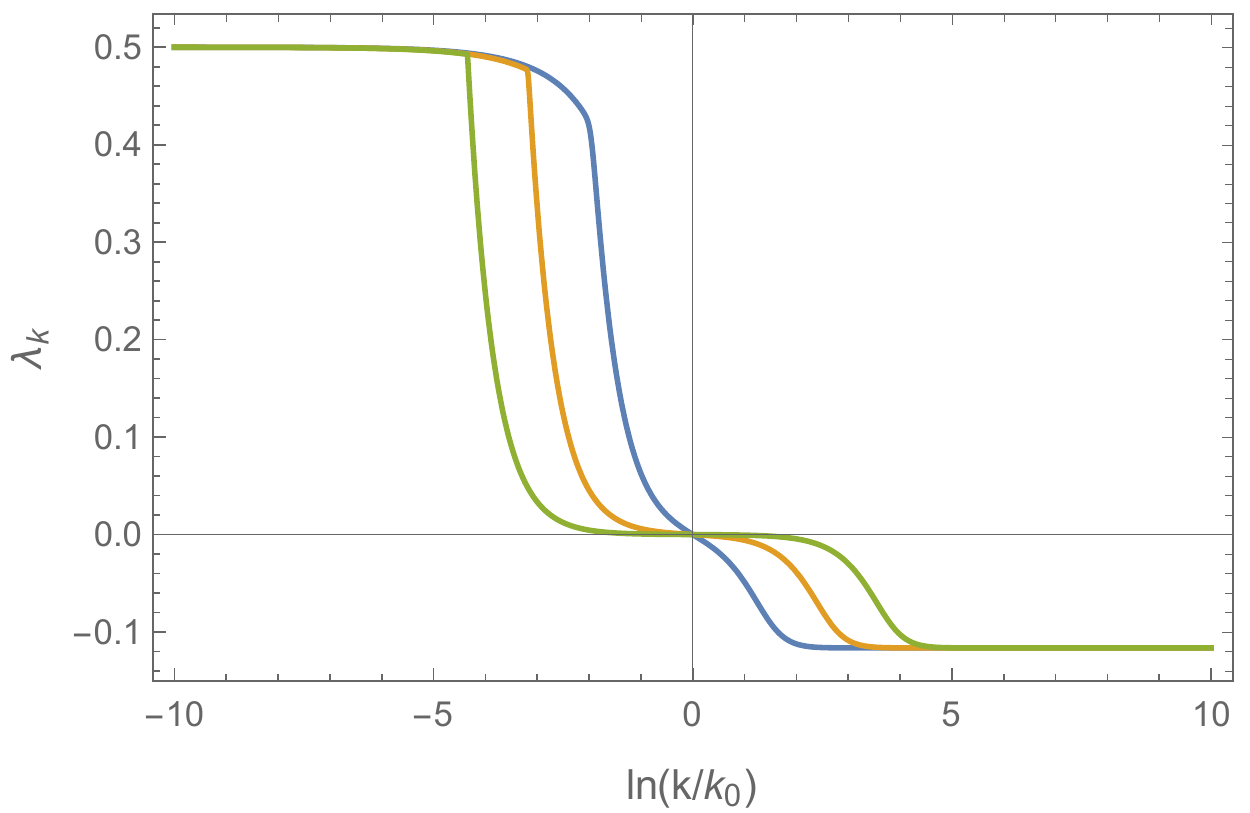} \; 
	\includegraphics[width = 0.48\textwidth]{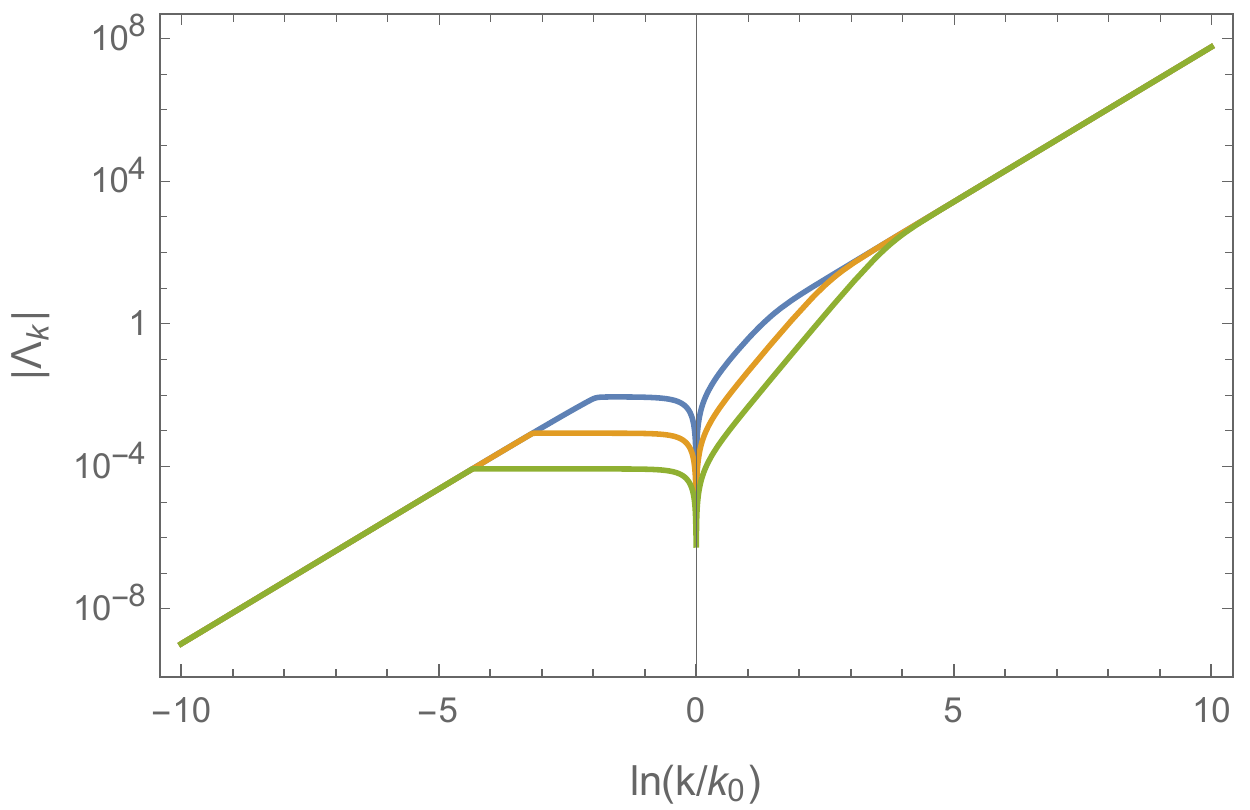} \\
	\caption{\label{FlowsIIIa} Illustration of typical RG trajectories situated between the  solid black lines shown in Fig.\ \ref{FlowDiaFEH} as a function of RG time $t$. The trajectories are obtained by solving the beta function obtained from the $p_0$-projection numerically with high precision. Starting from large values of $t$ and going towards lower coarse-graining scales, the trajectories exhibit three distinct scaling regimes: for $t \gg 1$ the running is controlled by the NGFP$_1$ indicating that the dimensionless couplings $g_k$ and $\lambda_k$ take their fixed point values. Lowering $t$, there is a cross-over to the GFP. This phase is characterized by the dimensionful couplings $G_k$ and $\Lambda_k$ being constant. For even lower values of $t$ the trjectories leave the scaling regime of the GFP and approach the IR-FP. Here the scale-dependence of $g_k$ and $\lambda_k$ follows the relation \eqref{IR-FP-scaling}. This implies that $\lim_{k \rightarrow 0} \Lambda_k = 0$ for all trajectories situated in this region. }
\end{figure}

  At this stage, it is interesting to investigate the RG trajectories located between the solid black lines in more detail. A set of example solutions situated in this region is shown in Fig.\ \ref{FlowsIIIa}. All trajectories are complete: the high-energy completion is provided by the NGFP$_1$. Lowering $k$ the flow crosses over to a ``classical regime'' where the dimensionful couplings $G_k$ and $\Lambda_k$ are constant. In terms of the phase diagrams given in Fig.\ \ref{FlowDiaFEH}, this corresponds to the regime where the trajectories linger in the vicinity of the GFP. At even lower values of $k$ the trajectories are captured by the IR-FP and follow the scaling law \eqref{IR-FP-scaling}. This feature has profound consequences for the mass appearing in the graviton propagator: \emph{based on the flow diagram the renormalized squared mass $\mu^2$ can never be negative}. The scaling induced by the IR-FP indicates that any positive $\Lambda_k$ at $k > 0$ is quenched to $\Lambda_0 = 0$ for trajectories in this region.

 \subsection{Fixed points of the non-relativistic system}
 \label{Sec.5.2}
 We proceed by analyzing the full system of beta functions including the flow of $\alpha_1$ and the dependence on the parameter $\alpha_2$. Before delving into the discussion, we make the following observation: if $\alpha_1 \not = 1$ or $\alpha_2 \not = 1$, the action \eqref{actionansatz} is no longer invariant with respect to the full diffeomorphism group Diff($\cM$). Instead the symmetry group is reduced to foliation-preserving diffeomorphisms Diff($\cM,\Sigma) \subset$ Diff($\cM$). This results in a mismatch between the dependence of the fields on spacetime points, $N(\tau,y)$ and the time-component of the vector \eqref{vdec} generating infinitesimal foliation-preserving diffeomorphism which depends on $\tau$ only (see \cite{Rechenberger:2012dt} for detailed discussions).\footnote{Projective Ho\v{r}ava-Lifshitz gravity \cite{Horava:2009uw} resolves this mismatch by restricting the lapse function $N$ to depend on $\tau$ only.} The change in the symmetry group potentially entails consequences for the Faddeev-Popov procedure leading to the ghost action \eqref{ghostgen}: it is unclear whether the scalar ghost and anti-ghost should be treated as fully spacetime-dependent fields. Solving this problem is identical to finding a good gauge-fixing procedure for non-projectable Ho\v{r}ava-Lifshitz gravity. Resolving this issue is beyond the scope of this work. Instead we will analyze two versions of the flow equations: the first one includes the contribution of the scalar ghost sector ($\bar{c}c$) and a second version where these contributions are not included in the beta functions.   
 
 In order to close the beta functions, the analysis also requires an assumption about the free parameter $\alpha_2$. Motivated by the previous section, we first consider the case where $\alpha_2 = 1$. We briefly comment on other values for this parameter at the end of this section.

  \begin{table}[t!]
  	\renewcommand{\arraystretch}{1.2}
 	\begin{center}
 		\begin{tabular}{|c | c | c| c  c   c | c  c  c |  }
 			\hline
 			$\alpha_2$ & Fixed Points & Ghost Sector & \multicolumn{3}{c |}{Couplings} & \multicolumn{3}{c |}{Critical Exponents}\\	
 			\hline	
 			~&~& & $g_*$ & $\lambda_*$ & $\alpha_{1,*}$ &  $\theta_1$ & $\theta_2$ & $\theta_3$ \\ 
 			\hline
 			\multirow{1}{*}{1}&NGFP$_1$& with $\bar{c}c$ &   $1.92$ &  $ -0.25 $       & $ 0.71$  & $6.12$   & \multicolumn{2}{c |}{$-4.24\pm 5.91 i$           }\\
 			\multirow{1}{*}{1}&NGFP$_1$& without $\bar{c}c$ &   $1.21$ &  $ -0.14 $       & $ 0.73$  & $5.09$   & \multicolumn{2}{c |}{$\phantom{-}0.16\pm 4.97 i$          }\\
 			\hline
 			\multirow{1}{*}{$\alpha_1$}&NGFP$_2$& with $\bar{c}c$ &   $65.03$ &  $ -10.96 $       & $ 8.66$  & $10.19$   & \multicolumn{2}{c |}{$-1.16\pm 2.66 i$           }\\
 			\multirow{1}{*}{$\alpha_1$} & NGFP$_2$& without $\bar{c}c$ &  $43.54$ & $-13.71$  & $14.53$  & $13.80$    &    \multicolumn{2}{c |}{$\phantom{-}0.23\pm 3.85 i$           }        \\
 			\hline
 		\end{tabular}
 	\end{center}
 	\caption{Fixed point structure of the full beta functions \eqref{betafctsdef} including the flow of $\alpha_1$. For $\alpha_2 = 1$ (top block) and $\alpha_1 = \alpha_2$ (bottom block), the fixed point without ghost contributions are connected continuously to the ones with ghost contributions. The NGFP$_1$ and NGFP$_2$ are not continuously connected though, therefore justifying their different labels.}
 	\label{Table3}
 \end{table}
 \emph{Fixed Points.} We first investigate the generalization of the GFP in the enlarged truncation. Expanding the beta functions in powers of $g$, one finds
 \begin{equation}\label{exp1}
 	\begin{split}
 		\beta_g &= 2 g + \mathcal{O}[g^2]*f_g(\lambda, \alpha_1) \, , \\
 		\beta_\lambda & = -2 \lambda +  \mathcal{O}[g]*f_\lambda(\lambda, \alpha_1) \, , \\
 		\beta_{\alpha_1} &=   \mathcal{O}[g]*f_{\alpha_1}(\lambda, \alpha_1) \, .\\
 	\end{split}
 \end{equation}
 The first terms in $\beta_g$ and $\beta_\lambda$ are fixed by the classical mass dimensions of the couplings and $f_g$, $f_\lambda$ and $f_{\alpha_1}$ encode the leading corrections to the classical result. From the expansion \eqref{exp1} one readily verifies that all three beta functions vanish for $g=\lambda=0$. Hence the system admits \emph{a one-parameter family of GFPs}:
 \be\label{line-of-GFPs}
 \text{GFPs:} \qquad \left(g_*, \lambda_*, \alpha_{1,*} \right) = \left(0,0, \alpha_{1,*} \right) \, .  
 \ee
 The critical exponents of these fixed points are fixed by the mass dimension of the couplings. Evaluating the stability matrix based on the expansion \eqref{exp1} one finds $\theta_1 = 2$, $\theta_2 = -2$, and $\theta_3 = 0$, with the latter corresponding to a marginal direction.
 
 In addition to the GFPs, the full system also possess a NGFP generalizing the fixed point NGFP$_1$ of the foliated Einstein-Hilbert truncation. Its position and stability properties are summarized in Table \ref{Table3} which gives the results for both cases where the scalar ghost contribution has been taken into account (labeled by ``with $\bar{c}c$'') and left out (labeled by ``without $\bar{c}c$''). We checked that the two fixed points are related by a continuous deformation when turning on the scalar ghost contribution in the beta functions. We note that the fixed point is situated close to but not within the surface $\alpha_1=1$ which is spanned by the foliated Einstein-Hilbert truncation. The position is rather insensitive to whether the scalar ghost contribution is included. The latter has a profound consequence on its stability properties though: in the absence of the $\bar{c}c$-contribution NGFP$_1$ is UV-attractive in all three directions. Upon including the scalar contribution in the beta function the fixed point turns into a saddle point with one UV-attractive and two UV-repulsive directions. 
 
 The occurrence of a line of GFPs in truncations of the form \eqref{actionansatz} has already been observed in \cite{Contillo:2013fua}. The non-relativistic viewpoint adopted in the present work offers a natural explanation of this phenomenon. Let us set $\Lambda_k = 0$ for the time being, i.e., all component fields are taken to be massless. Inspecting the dispersion relations of the component fields, we encounter the generic form
 \be\label{disp1}
 \Gamma_k^{(ij)} \propto a_i \, p_0^2 + b_i \, \vec{p}^{\,2} \, , \qquad c_i^2 \equiv \frac{b_i}{a_i} \, . 
 \ee
 The explicit values for the coefficients $a_i$ and $b_i$ can be read off from Table \ref{table.hessian} and $c_i$ is the resulting ``speed of light'' associated with a given component field. For instance, the transverse-traceless modes $h_{ij}$ and the scalar $B$ propagate with
 \be\label{disp1exp}
 c_{h}^2 = \frac{1}{\alpha_1} \, , \quad \text{and} \qquad c_{B}^2 = 1 + 2 \alpha_1 - 2 \alpha_2 \, , 
 \ee
 respectively. \footnote{In the scalar sector, comprising the fields $\Psi$ and $\hat{N}$ the dispersion relations do not have the simple structure \eqref{disp1} owed to the off-diagonal terms in the propagator. Generically the eigenvalues of this matrix involve square-roots. In the special case $\alpha_1 = \alpha_2 = 1$, all equations reduce to their relativistic form and one has $c_i = 1$ for all component fields.}
 Keeping $\alpha_2$ fixed, the system then admits a one-parameter family of non-interacting theories, where the component fields satisfy different dispersion relations controlled by the value of $\alpha_1$. This feature then generates the line of GFPs \eqref{line-of-GFPs}. The fact that differences in propagation speeds are observable quantities suggests that each point on the line actually corresponds to a physically distinct theory. This also entails that lines of GFPs should be a rather generic property of non-relativistic systems where one has several fields with distinct dispersion relations. 
 
 Table \ref{Table3} also reports a NGFP found within the alternative identification $\alpha_1 = \alpha_2$. This NGFP comes with identical stability properties as the $\alpha_2 = 1$ case, albeit at much larger values of the couplings. Again one readily checks that the NGFPs found with and without the scalar ghost contributions are connected by analytic continuation. While the positions and stability coefficients suggest that the fixed points for $\alpha_2=1$ and $\alpha_1 = \alpha_2$ are also continuously connected we have not been able to construct such a deformation explicitly. Building on the interpolation $\alpha_2 = 1 + \gamma(\alpha_2 - 1)$, $\gamma \in [0,1]$, the lines followed by the fixed point as a function of $\gamma$ terminate around $\gamma \approx 0.7$. We conclude that the NGFP$_1$ is robust under a wide range of values for $\alpha_2$. Moreover, it is the scalar ghost sector which has an decisive effect on the stability properties of the fixed point.

\section{Summary and discussion}
\label{sect.conc}
 The presence of a foliation on spacetime is essential for transiting from quantum gravity formulated on Euclidean signature spacetime to its Lorentzian counterpart. One way to implement this structure is through the Arnowitt-Deser-Misner (ADM)-decomposition of the metric degrees of freedom. In this work, we studied the RG flows arising within this setting using non-perturbative functional renormalization group methods. Our work reports the first fluctuation field computation in the ADM-framework, studying the RG flow of the graviton two-point function. It complements earlier investigations based on the ADM-formalism \cite{Manrique2011,Rechenberger:2012dt,Contillo:2013fua,Biemans:2016rvp,Biemans:2017zca,Houthoff:2017oam} and covariant computations utilizing a foliation gauge fixing \cite{Knorr:2018fdu,Eichhorn:2019ybe,Knorr:2022mvn}. It is also closely related to the Lorentzian signature computations based on the Wetterich equation \cite{Banerjee:2022xvi,DAngelo:2022vsh,DAngelo:2023tis}. At the technical level, these computations are significantly more complex than their counterparts in the covariant setting reviewed in \cite{Pawlowski:2020qer}. This is owed to the increased number of component fields carrying the gravitational degrees of freedom as well as the  proliferation of potential vertex structures coming with independent couplings.
 
 We give a comprehensive introduction to the general setup and structures featuring in these computations. In particular, we highlight the differences between fluctuation computations in the covariant and foliated frameworks. The fact that the latter naturally also comprises non-relativistic theories like Ho\v{r}ava-Lifshitz gravity \cite{Horava:2009uw} introduces new elements. In particular, one finds that, generically, projections onto tensor structures involving the spatial and time-parts of the fluctuation field momentum may lead to different results. Conceptually, this implies that one is dealing with different coupling constants of the theory which, in the covariant setting, exhibit the same flow due to the enhanced symmetry.
 
 In order to illustrate the general framework, we performed an explicit computation of the gravitational RG flow projected onto the graviton two-point function. The setting is based on a truncation of the effective average action containing all two-derivative terms compatible with foliation-preserving diffeomorphisms, cf.\ eq.\ \eqref{actionansatz}. This action is used to derive the graviton two-point function as well as the 3- and 4-point vertices required to close the flow equation. By projecting the RG equation onto the 2-point function, we derive the beta functions for Newton's coupling, the cosmological constant (featuring as a graviton mass term in the 2-point function) as well as one coupling $\alpha_1$ encoding deviations from general relativity. Furthermore, the results depend parametrically on a second Lorentz-symmetry violating coupling $\alpha_2$ which does not enter in the 2-point function under consideration but appears on the right-hand side of the flow equation.
 
 The fixed point structure and phase diagram resulting from the foliated Einstein-Hilbert action is discussed in detail. As a key result, all projections studied in this work identify a non-Gaussian fixed point (NGFP) suitable for asymptotic safety.  It serves as a UV-attractor for the two couplings retained within the projection. This puts the existence of the NGFP already identified in background field computations on spherical \cite{Manrique:2011jc,Rechenberger:2012dt}, toroidal \cite{Houthoff:2017oam}, and cosmological backgrounds \cite{Biemans:2016rvp,Platania:2017djo} on firm grounds. The NGFP seen in our work is part of an intricate phase diagram, see Fig.\ \ref{FlowDiaFEH}. In the physically relevant region, its structure is governed by the interplay of the GFP, the NGFP, and an IR-FP. The latter controls the IR-behavior of RG-trajectories which, in principle, could lead to a negative squared graviton mass $\mu^2$. The fixed point quenches this mass to zero. This restricts the values of the squared graviton mass supported by the phase diagram to $\mu^2 \ge 0$.
 
 Remarkably, our phase diagram is qualitatively identical to the one derived from similar fluctuation field computations in the covariant setting \cite{Christiansen:2012rx,Christiansen:2014raa}.\footnote{The phase diagram is also very similar to the one found for the Einstein-Hilbert truncation at the background level, first constructed in \cite{Reuter:2001ag}. In this case, the IR-FP is replaced by a singular line which leads to a termination of the RG-trajectories at finite values of the coarse-graining scale though.} This is rather striking, since the underlying computations are based on an entirely different implementation of the gravitational degrees of freedom. Moreover, the vertex structures used in the right-hand side of the flow equation are manifestly different since the relation between the covariant fluctuation field and the fluctuation fields in the ADM-formalism are non-linear. Taken together, this indicates a remarkable stability of the phase diagram related to the graviton 2-point function.  
 
 The space of action functionals which are invariant with respect to diffeomorphisms spans a subspace of all action functionals compatible with foliation preserving diffeomorphisms. The enhanced symmetry implies that RG flows on the former are closed \cite{Rechenberger:2012dt,Knorr:2018fdu}. The Wetterich equation adapted to the ADM-formalism allows to study RG flows on the larger theory space as well. In this work, we did this by deriving the beta function for the coupling $\alpha_1$ encoding different speeds of light for the component fields in the construction. We show that all projections within this truncation possess a NGFP which is the analogue of the NGFP seen in the foliated Einstein-Hilbert case. Based on symmetry arguments, this NGFP should be located on the surface with enhanced symmetry, $\alpha_1 = \alpha_2 = 1$. Since the Wetterich equation for the ADM-formalism contains diffeomorphism breaking terms the fixed point is slightly shifted away from this surface though. The stability properties of the NGFP depend on the implementation of the scalar ghost sector. In case that the scalar ghost is left out (as suggested by the reduced symmetry of the setup) one finds that the NGFP is a UV-attractor on the projection space. This is in agreement with the background computation \cite{Contillo:2013fua}, which identified a similar UV-attractive fixed point once diffeomorphism-breaking couplings are included.
 
 Our work also makes a notable step forward in connecting continuum computations and Monte Carlo simulations of the gravitational path integral. In particular, the framework introduced in our work gives access to the spatially averaged correlation functions accessible in the Causal Dynamical Triangulations (CDT) program \cite{Ambjorn:2012jv,Loll:2019rdj}. In this context, it is important to highlight that the phase diagram obtained in the present work differs from the one obtained at the background level \cite{Reuter:2001ag,Manrique:2011jc} by the addition of an IR-fixed point. This feature may be interesting when interpreting the RG flows \cite{Ambjorn:2014gsa,Ambjorn:2020rcn} obtained from Monte Carlo simulations in the context of CDT. Both settings share the same foundations in the sense that they build on a foliation structure and admit anisotropy parameters distinguishing between space and time. 
 We hope to come back to this intriguing possibility in the future. 
 
Obviously, it will also be interesting to investigate the robustness of the fixed point structure and phase diagram shown in Fig.\ \ref{FlowDiaFEH} once matter degrees of freedom are included. Specifically, one may want to ad massless gauge fields supplemented by Lorentz symmetry breaking parameters, similar to $\alpha_1$ and $\alpha_2$ introduced in the gravitational sector. This setting may allow to study differences in the light cone structure for the two fields which can be constrained by multi-messenger astronomy. Moreover, the framework set up in the present work allows for the direct transition to Lorentzian signature computations. Results along these lines will be reported in \cite{WangInPrep1} and \cite{WangInPrep2}, respectively.
\section*{Acknowledgments}
We thank J.\ Ambj{\o}rn, M.\ Becker, A.\ Bonanno, G.\ P.\ de Brito, A.\ Ferreiro, B.\ Knorr, G.\ Korver, A.\ Koshelev, A.\ Pereira, M.\ Reuter, and M.\ Schiffer for discussion and insightful comments on the manuscript. JW acknowledges the China Scholarship Council (CSC) for financial support.

\appendix
\section{Technical Annex}
\label{App.A}
The derivation of the results presented in Sect.\ \ref{Sec.3.1} is technically quite involved. This appendix collects some additional background information and lengthy formulas underlying these computations. We start by briefly outlining the implementation of a foliation structure within {\tt Mathematica} in App.\ \ref{App.A1}. App.\ \ref{App.A2} collects the propagators for the component fields and the actions underlying the derivation of the gravity-ghost vertices are given in App.\ \ref{App.D}.
\subsection{Implementation within Mathematica}
\label{App.A1}
In the ADM-formalism the relations \eqref{gtoadm} decompose the spacetime $g_{\mu\nu}$ into the lapse function $N$, a shift vector $N_i$, and spatial metric $\sigma_{ij}$. As a result, the four-dimensional spacetime $\cM$ inherits a foliation structure 
\be\label{Mdec}
\cM = \mathbb{R} \times \Sigma_\tau \, . 
\ee
The dynamics of the theory is expressed in terms of time-derivative $\partial_\tau$ and covariant derivatives $D_i$ constructed from the metric $\sigma_{ij}$ on the spatial slices.  

In order to implement the decomposition \eqref{Mdec} in the {\tt xAct}-package \cite{xactref,Nutma:2013zea} for {\tt Mathematica}, we define the three-dimensional manifold $\Sigma_\tau$ and a one-dimensional manifold $\mathbb{R}$. These correspond to the two factors in \eqref{Mdec}, respectively. Subsequently, the code defines the four-dimensional spacetime $\mathcal{M}$ as the product of the two factors. The field content is then coded as the spatial metric $\sigma_{ij}(\tau,y)$, one scalar field $N(\tau,y)$, and one vector field $N_i(\tau,y)$. The indices are taken to be covariant with respect to $\Sigma_\tau$.  The time-derivative and spatially covariant derivatives are defined as the covariant derivatives on $\mathbb{R}$ and $\Sigma_\tau$, respectively. They are compatible with the metrics on the two submanifolds, respectively. 

Building on this setup, all propagators and vertices needed in our computation can be generated via the {\tt Perturbation} command provided by the {\tt xAct} package. The expressions for the vertices obtained in this way are very lengthy though. This holds specifically for the $3$- and $4$-point vertices which contain hundreds of contributions. Therefore, we will not give the explicit form of these contributions within this paper. The {\tt mathematica}-code generating them is available on request.

\subsection{Propagators for the component fields}
\label{App.A2}

The $2$-point functions including the contribution from the gauge fixing term have been summarized in Table \ref{table.hessian}. The propagators required in the evaluation of eq.\ \eqref{twopointproj} are obtained as the inverse of this matrix. Defining the propagator matrix including the contributions of the scalar regulator $\cR_k(p^2)$,
\be\label{def.propmat}
\cG \equiv \left( \Gamma_k^{(2)} + \cR_k\right)^{-1} \, , 
\ee
we list the matrix elements in Table \ref{table.prop}. In addition to the definition of the $4$-momentum, $p^2 = p_0^2+\vec{p}^{\,2}$, the table uses the shorthand notations 
\begin{equation}\label{dispersionsshort}
	\begin{split}
		f_a =  \alpha_1 p_0^2 +\vec{p}^{\,2} \, , \;
		f_b =\frac{1}{\alpha_1} p_0^2+ \vec{p}^{\,2} \, , \;
		f_d = \alpha_2 p_0^2 +\vec{p}^{\,2} \, , \;
		f_e  =\frac{1}{\alpha_2} p_0^2 + \vec{p}^{\,2} \, ,
	\end{split}
\end{equation}
for capturing the $\alpha$-dependence of the dispersion relations.
\begin{table}[t!]
	\renewcommand{\arraystretch}{1.5}
	\begin{center}
		\begin{tabular}{ll}
			\hline \hline
			fields $(i,j)$  & $\cG^{(ij)}_{k}$ \\ \hline \hline
			$h_{ij} h^{kl}$ & $32 \pi G_k \, \frac{1}{f_a - 2 \Lambda_k + R_k}\, {\Pi_h}^{ij}_{~~kl} $ \\  
			$v_i v^j$ & 
			$16 \pi G_k \, \frac{1}{ f_a -2 \Lambda_k + R_k} \, {\Pi_u}{}^i{}_j$ \\
			$EE$ & 
			$48 \pi G_k \, \frac{1}{f_a -2  \Lambda_k +R_k}$ \\
				$u^i u_j$ & 
			$\frac{16 \pi G_k}{\alpha_1}  \, \frac{1}{ f_b+ R_k} \, {\Pi_u}{}^i{}_j $ 	\\ 
			$BB$ & 
			$16 \pi G_k \,  \frac{1}{ p^2 +2 \alpha_1 f_b -2 \alpha_2 f_e + (2 \alpha_1-2\alpha_2 +1) \, R_k}$ \\ \hline
			$\Psi\Psi$ & 
			$96 \pi G_k \Big(\frac{p^2 +R_k}{f_a p^2 -3 p^2 f_d + 7 p^2 \Lambda_k - 6 \Lambda_k^2+(f_a- 2 p^2 - 3 f_d+7\Lambda_k)R_k - 2 R_k^2}\Big)$ \\
			$\hat{N}\hat{N}$ & 
			$- 8 \pi G_k \Big( \frac{6 f_d -2 f_a-3 p^2 -2 \Lambda_k +R_k}{f_a p^2 -3 p^2 f_d + 7 p^2 \Lambda_k - 6 \Lambda_k^2+(f_a- 2 p^2 - 3 f_d+7\Lambda_k)R_k - 2 R_k^2)}\Big)$ 	\\ 
			$\Psi \hat{N}$ & 
			$48 \pi G_k \Big( \frac{p^2 + R_k -2 \Lambda_k}{f_a p^2 -3 p^2 f_d + 7 p^2 \Lambda_k - 6 \Lambda_k^2+(f_a- 2 p^2 - 3 f_d+7\Lambda_k)R_k - 2 R_k^2}\Big)$ 
		 \\ \hline
			$\bar{c}c$ & $\frac{1}{\sqrt{2} \, p^2}$ \\
			$\bar{b}^i b_i$ & $\frac{1}{\sqrt{2} \, p^2} \, \Pi_u{}_j{}^i$
			\\ \hline \hline
		\end{tabular}
	\end{center}
	\caption{\label{table.prop} Matrix elements of the regulated propagators required in the evaluation of \eqref{twopointproj}. For the diagonal blocks given in Table \ref{table.hessian}, the inversion is straightforward and the results are given in the first block. For the scalar sector spanned by the fields $\Psi$ and $\hat{N}$ the propagator is found by inverting a $2\times 2$-matrix. The determinant arising from this inversion leads to the rather complicated dispersion relations listed in the second block. The propagators in the ghost sector are given in the third block.}
\end{table}
The projectors $\Pi$ carrying the index structure of the propagators are defined in eqs.\ \eqref{projectors1} and \eqref{projectors2}.

\subsection{Ghost actions on a foliated spacetime}
\label{App.D}
We close the technical annex by giving the ghost action resulting from evaluating the general formula \eqref{ghostgen} for the specific choice of gauge fixing functional \eqref{gaugechoice}. The result is used to extract the propagators and interaction vertices involving the scalar (anti-)ghost $\bar{c}, c$ and the vector (anti-)ghosts $\bar{b}^i, b_j$, respectively. In order to keep the complexity at a manageable level we chose the background to be flat space. Moreover, all derivatives act on the right, $\partial_\tau N_i c =N_i (\partial_\tau c) +(\partial_\tau N_i) c$. The contribution proportional to the scalar anti-ghost $\bar{c}$, arising from the first term in \eqref{ghostgen} is
\begin{equation}
	\begin{split}
		\Gamma^{\text{scalar}}_{\text{ghost}}  = -\sqrt{2} \, \int d\tau d^dy \, \bar{c} \Big[ &
		\partial_\tau^2 c N 
		+ \partial_\tau b^k \partial_k N 
		- \partial_\tau N N^i \partial_i c 
		+ \partial^i \partial_\tau N_i c 
		+ \partial^i b^k \partial_k N_i \\
		& + \partial^i N_k \partial_i b^k 
		+ \partial^i \sigma_{ki} \partial_\tau b^k 
		+ \partial^i N_k N^k \partial_i c 
		+ \partial^i N^2 \partial_i c \\ & - \frac{1}{2} \partial_\tau \sigma^{ij} c \partial_\tau \sigma_{ij}
		- \frac{1}{2}  \partial_\tau \sigma^{ij} b^k \partial_k \sigma_{ij} - \partial_\tau \partial_k b^k - \partial_\tau N^i \partial_i c \Big]\, . 
	\end{split}
\end{equation}
In the vector sector, the terms proportional to the anti-ghost $\bar{b}^i$ are
\begin{equation}
	\begin{split}
		\Gamma^{\text{vec}}_{\text{ghost}}  =  -\sqrt{2}  \, \int  \, d\tau & d^dy \, \bar{b}^i \Big[  
		 \partial^2_\tau N_i c 
		 +  \partial_\tau b^k \partial_k N_i 
		 +  \partial_\tau N_k \partial_i b^k 
		 +  \partial_\tau \sigma_{ki} \partial_\tau b^k \\
		 &
		 +  \partial_\tau N_k N^k \partial_i c +  \partial_\tau N^2 \partial_i c 
		-  \partial_i \partial_\tau c N 
		-  \partial_i b^k \partial_k N \\
		&
		 +  \partial_i N N^k \partial_k c  - \frac{1}{2}   \partial_i \sigma^{mn} c \partial_\tau \sigma_{mn}-\frac{1}{2}  \partial_i \sigma^{mn} b^k \partial_k \sigma_{mn}
		- \partial_i \partial_k b^k \\
		&
		- \partial_i N^m \partial_m c 
		+ \partial^j c \partial_\tau \sigma_{ij} 
		+  \partial^j b^k \partial_k \sigma_{ij} +  \partial^j \sigma_{jk} \partial_i b^k 
		+  \partial^j \sigma_{ik} \partial_j b^k \\ &
		+ \partial^j N_j \partial_i c 
		 +  \partial^j N_i \partial_j c \, \Big] . 
	\end{split}
\end{equation}

\section{Trace computations}
\label{App.B}
The beta functions given in Sect.\ \ref{Sec.3.2n} are obtained by projecting the Wetterich equation \eqref{WetterichEq} onto the $2$-point function of the transverse-traceless fluctuation field $h_{ij}(x)$. Their computation requires evaluating the traces appearing on the right-hand side of eq.\ \eqref{twopointproj}. In a flat background, this is conveniently done by using standard momentum space methods for evaluating Feynman diagrams. In this appendix, we provide some general identities useful in these computations in App.\ \ref{App.B1}. The computation of a typical diagram built from a $4$-point vertex and $3$-point vertices is illustrated in App.\ \ref{App.B2}. We also briefly discusses the analytic properties of the loop integrals.
\subsection{Structural aspects of the traces}
\label{App.B1}
The right-hand side of the projected flow equation contains one-loop diagrams built from a pair of $3$-point vertices (bubble diagrams) and a $4$-point vertex (tadpole diagrams)
\begin{equation}
	\begin{split}
		T_3 = \, &-\frac{1}{2} \text{STr} [\mathcal{G}_{\chi_{a^i} \chi_{a^j}} {\Gamma}^{(h h \chi_{a^j} \chi_{a^k})} \mathcal{G}_{\chi_{a^k} \chi_{a^l}}{ \partial_t R_k}^{\chi_{a^l} \chi_{a^i}}]\\
		T_4 = \, & \text{STr} [\mathcal{G}_{\chi_{a^i} \chi_{a^j} } \Gamma^{({h  \chi_{a^j} \chi_{a^k}})} \mathcal{G}_{\chi_{a^k} \chi_{a^l}}  \Gamma^{( h  \chi_{a^l} \chi_{a^m})} \mathcal{G}_{\chi_{a^m} \chi_{a^n} }{ \partial_t R_k}^{\chi_{a^n} \chi_{a^i}}] \, . 
	\end{split}
\end{equation}
Here the subscript on $T$ refers to the vertices contained in the trace. The trace contains a sum over all fluctuation fields which can propagate in the regularized loop and subsequently we will use superscripts to single out the contributions of specific fields to the complete trace. In addition, the trace also contains an integral over loop momenta $q_\mu = (q_0, q_i)$. The propagators and vertices depend on both the loop momentum and the external momentum associated with the transverse-traceless field $h_{ij}(p_0, \vec{p})$. The integrals over the spatial loop momenta have the generic form
\begin{equation}\label{spatialintegralmaster}
I_n \equiv \int d^3 q \, f(\vec{q}^{~2}) \, T^{i_1 i_2 \cdots i_n} q_{i_1} q_{i_2} \cdots q_{i_n} \, . 
\end{equation}
The $T^{i_1 i_2 \cdots i_n}$ are tensors constructed from the fields and external momenta allowed by our projection. Canonical building blocks are, for example, $h^{ij}(p_0,\vec{p})$ and $p^i$. The parts depending on the magnitude of the spatial loop momentum have been factored out and collected in the function $f(\vec{q}^{\,2})$. 

In the present computation, the maximum number of loop momenta contracted with  $T^{i_1 i_2 \cdots i_n}$ turns out to be $n=6$. Higher-order terms are outside of the projection subspace and do not contribute to the computation. Following \cite{Peskin:1995ev}, the integrals \eqref{spatialintegralmaster} can be simplified as follows. For $n$ odd the contributions vanish due to the anti-symmetry in sending $q_i \mapsto - q_i$. For even $n$ the integrands can be reduced to functions depending on $\vec{q}^{\,2}$ only. For the integrals defined in eq.\ \eqref{spatialintegralmaster}, the relevant identities are
\begin{equation}\label{intids1}
	\begin{split}
		I_2 & =  \frac{1}{3} \, \int d^3 q f(\vec{q}^{\,2}) T_{ij} \, \delta^{ij} \vec{q}^{\,2} \, ,\\
	I_4  & =  \frac{1}{15} \int d^3 q f(\vec{q}^{\,2}) T_{ijkl} (\delta^{ij}\delta^{kl}+\delta^{ik}\delta^{jl}+\delta^{il}\delta^{jk}) (\vec{q}^{\,2})^2,\\
	I_6  & = \frac{1}{105} \int d^3 q f(\vec{q}^{\,2}) T_{ijklmn}  \left( \delta^{ij} \delta^{kl} \delta^{mn} + 14 \; \text{permutations} 
	%
		\right) (\vec{q}^{\,2})^3 \, . 
	\end{split}
\end{equation}

The external legs associated with the transverse-traceless field $h_{ij}$ have to be attached to the $3$- and $4$-point vertices. In terms of computational efficiency, it is convenient to retain these fields in the intermediate steps in order to deal with scalar instead of tensorial quantities. Upon carrying out the integration over loop momenta, they can be stripped. The projection onto the tensor structure $\Pi_h$ appearing on the left-hand side is then evoked by contracting the resulting matrix with the projectors $\Pi_h$ defined in eq.\ \eqref{projectors1}. The coefficients $T_{p_0}$, $T_{\vec{p}}$, and $T_0$ defined in \eqref{projectionrhs} are the read off by matching powers of the external momenta. 

\subsection{Computations of selected traces}
\label{App.B2}
In general, the evaluation of the tadpole diagrams is simpler than the one of the bubble diagrams. Hence we start with an example based on the $4$-point vertex containing four gravitons. Subsequently, we discuss the projection of the traces associated with bubble-diagrams, expanding the trace arguments in terms of the external momenta.
\subsubsection{The tadpole diagram containing the $4$-graviton vertex}
The contribution originating from the $4$-graviton vertex has the form
\begin{equation}
 T_4^{hh} =	-\frac{1}{2} \text{Tr}[ \mathcal{G}_{hh} \Gamma^{(hhhh)}_{k} \mathcal{G}_{hh} {\partial_t \cR_k}^{hh}] \, .
\end{equation}
The propagators are listed in Table \ref{table.prop} and the $4$-point vertex is generated from the computer algebra. The trace contains an integration over loop-momentum. The explicit evaluation of these integrals requires choosing a regulator function $R_k$. In practice, we have opted for a Litim-type regulator \cite{Litim:2000ci,Litim:2001up} where $R_k(q^2)  = (k^2 - q^2)\Theta(k^2 - q^2)$. In this case, the integration is restricted to the domain where $q_0^2 + \vec{q}^{\,2} \le k^2$. Performing all index contractions and using the identities \eqref{intids1},  one finds
\begin{equation}\label{traceexample1}
	\begin{split}
	T_4^{hh} =	- \int_{q^2 \le k^2}  & \frac{dq_0 d^3 \vec{q}}{(2 \pi)^4} \bigg[  \frac{(k^2( \eta_N-2)-q^2 \eta_N)(5\vec{q}^{~2}+ 9 \alpha_1 q_0^2-4 \alpha_2 q^2_0 - 14 k^2 \lambda_k)}{5(q_0^2(\alpha_1 -1 )+k^2(1-2\lambda_k))^2} \\
		& + \frac{(k^2 (\eta_N-2) - q^2 \eta_N)}{(q_0^2(\alpha_1 -1)+k^2(1-2\lambda_k))^2}  \left( \frac{9 \alpha_1 -4 \alpha_2}{5} p_0^2 + \vec{p}^{\,2} \right) \bigg] h^{ij}(-p) h_{ij}(p) \, . \\
	\end{split}
\end{equation}

This expression illustrates the following feature. We have picked the regulator to be a function of $q^2 =q_0^2 + \vec{q}^{\,2}$. The propagators entering our computation  contain the loop momentum $q_0^2$ and $\vec{q}^{\,2}$ in different linear combination. In eq.\ \eqref{traceexample1} this manifests itself in form of the terms $q_0^2(\alpha_1-1)$ in the denominators. As a consequence the integrand may exhibit poles when performing the integral over $q_0$. However, we find that there are regions in the parameter space of $\alpha_1$ and $\lambda_k$ where the poles are absent and the integral is finite. For the integrands in eq.\ \eqref{traceexample1} this requires that the following range of couplings should be excluded:
\be\label{hhbound}
hh: \qquad -1 \le \frac{1-2 \lambda_k }{\alpha_1-1} \le 0 \, . 
\ee
Note that this bound is satisfied by the foliated Einstein-Hilbert truncation where $\alpha_1 = \alpha_2 = 1$ and $q_0$ drops out of the denominators. 

We proceed by evaluating \eqref{traceexample1}. Since $\vec{q}$ appears in the numerator only, performing this integration is rather straightforward. Subsequently, we integrate over $q_0$. Assuming the absence of poles, this integration results in inverse trigonometric functions. Since the outcome is rather lengthy, we illustrate the generic structure based on the coefficient multiplying the tensor structure  $h^{ij} h_{ij} ~\vec{p}^{~2}$,
\begin{equation}\label{trace4hh}
	\begin{split}
 T_4^{hh}  = 	 \Big( P_1(q_0) + P_2 (q_0)~f^{hh}_1(q_0)  + P_3(q_0)~ f^{hh}_2(q_0) \left. \Big) \right|^{q_0=k}_{q_0=-k} \times
h^{ij} h_{ij} ~\vec{p}^{\,2}  \, . 
\end{split}
\end{equation}
The trigonometric contributions are captured by
\be\label{trigfcts}
f^{hh}_1(q_0) =  \text{arctan} \left( \frac{q_0}{\sqrt{k^2-q_0^2}} \right) \, , \quad f^{hh}_2(q_0) =  \text{arctanh} \left( \frac{q_0}{\sqrt{k^2-q_0^2}} \sqrt{\frac{\alpha_1-2\lambda_k}{2\lambda_k-1}} \right) \, .
\ee
The polynomials appearing in this expression read 
\begin{equation}
	\begin{split}
		P_1(q_0)=& - \frac{q_0 \sqrt{k^2-q_0^2}}{60 \pi ^3 \left(\alpha_1-1\right){}^2 \left(1-2 \lambda_k\right) \left(k^2 \left(1-2 \lambda _k\right)+\left(\alpha_1-1\right) q_0^2\right)}\\
		& \Big(\alpha_1 k^2 \left(\lambda _k \left(10-4 \eta _N\right)+5\right)+k^2 \left(-10 \lambda _k+8 \lambda _k^2 \eta _N-4 \lambda _k \eta _N+\eta _N\right) \\ & \qquad + \alpha_1^2 k^2 \left(\eta _N-5\right)+\left(2 \lambda _k-1\right) \eta _N q_0^2+\alpha_1 \left(1-2 \lambda _k\right) \eta _N q_0^2\Big) \\
		P_2(q_0)=& \frac{k^2}{60 \pi ^3 \left(\alpha_1-1\right){}^3} \times
		\left(8 \lambda _k \eta _N-5 \alpha_1 \left(\eta _N-2\right)+\eta _N-10\right),\\
		P_3(q_0)=& -\frac{k^2 \sqrt{\alpha_1-2 \lambda _k}}{60 \pi ^3 \left(\alpha_1-1\right){}^3 \left(2 \lambda _k-1\right){}^{3/2}} \times \Big(\alpha_1 \left(6 \lambda _k \eta _N-5 \eta _N+20-20 \lambda _k\right) \\  & \qquad -16 \lambda _k^2 \eta _N 
		 +10 \lambda _k \left(\eta _N+2\right)+\alpha_1^2 \left(\eta _N-5\right)-15\Big).
	\end{split}
\end{equation}
Matching the result \eqref{trace4hh} to the definition \eqref{projectionrhs}, one then obtains the contribution of this specific tadpole diagram to the beta functions.

\begin{figure}[t]
	\begin{center}
		\includegraphics[width=0.48\textwidth]{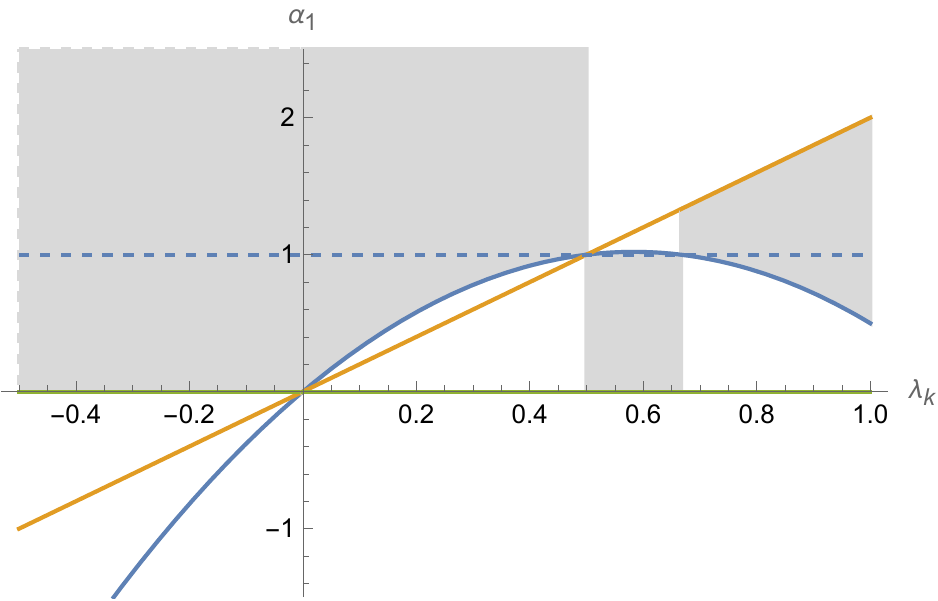}
		\includegraphics[width=0.48\textwidth]{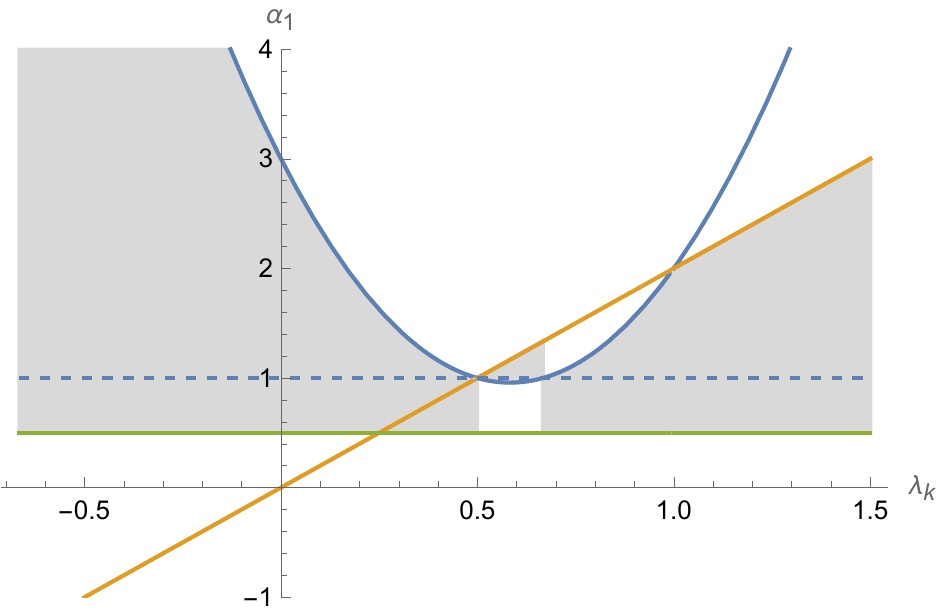}
		\caption{\label{Range2} Illustration of the inequalities \eqref{hhbound} and \eqref{otherbounds} for $\alpha_2 = 1$ (left panel) and $\alpha_1 = \alpha_2$ (right panel). The inequalities are satisfied in the white regions, implying that the loop integrals do not converge for this set of couplings. The gray regions are admissible. Note that the foliated Einstein-Hilbert truncation $\alpha_1 = \alpha_2 = 1$ is situated in the gray region and all loop contributions are finite in this case.}
	\end{center}
\end{figure}

The remaining tadpole contributions can be computed along the same lines. Again, one encounters inverse trigonometric functions which put bounds on the space of admissible couplings. Since the component fields come with different dispersion relations, one actually finds a set of inequalities. Retaining both parameters $\alpha_1$ and $\alpha_2$ we find the following exclusion regions
\be\label{otherbounds}
\begin{split}
BB: \qquad & -1 \le \frac{2 \alpha_1 -2 \alpha_2+1}{2 (\alpha_2- \alpha_1)} \le 0 \, , \\
NN: \qquad & -1 \le - \frac{2 -7 \lambda_k + 6 \lambda_k^2}{2+ \alpha_1 - 3 \alpha_2} \le 0 \, , \\
uu: \qquad & -1 \le \frac{\alpha_1}{1-\alpha_1} \le 0 \, . 
\end{split}
\ee
For $\alpha_2 = 1$ the condition for $uu$ is a subset of $BB$, so that one has to consider only three inequalities in this case. The admissible values for the couplings leading to convergent loop integrals are most conveniently found graphically and are illustrated in Fig.\ \ref{Range2}. We note that the traces found for the foliated Einstein-Hilbert truncation are always convergent due to our choice of regulator.

\subsubsection{Bubble diagrams}
We complete the appendix with a brief discussion of the contributions resulting from bubble diagrams, encoded in the first trace of eq.\ \eqref{twopointproj}. We can always shift the loop momentum so that the argument of the regulator is $q^2$. The trace contributions then take the form
\begin{equation}\label{T3trace}
T_3 = 		\text{STr} \left[\mathcal{G}_{\chi_{a} \chi_{b}}(q) \Gamma^{(h  \chi_{b} \chi_{c})}_k(p,q)\mathcal{G}_{\chi_{c} \chi_{d}} (p+q) 
		\Gamma^{(h  \chi_{d} \chi_{e})}_k(p,q)\mathcal{G}_{\chi_{e} \chi_{f}}(q) \,  \partial_t \cR_k^{\chi_{f} \chi_{a}}(q) \right] \, ,
\end{equation}
where we have highlighted the momentum dependence of the building blocks.

The schematics of \eqref{T3trace} shows that there is always one propagator depending on both the internal and external momentum. This momentum dependence does not align with the step function contained in the regulator. In general, this would then lead to a quite complicated integration region. This complication can be avoided by noting that our computation does not need to track the full dependence of the trace on the external momentum. It is sufficient to extract the terms proportional to $\vec{p}^{\,2}$ and $p_0^2$ to extract the beta functions. Thus, we can expand the middle propagator in orders of external momenta and only keep the terms contributing to our projection. The relevant terms follow from a standard Taylor expansion in multiple variables
\begin{equation}
	\begin{split}
		f( |\vec{p}+\vec{q}|, |p_0+q_0|) = & f( |\vec{q}| , |q_0|) + \frac{ \partial  f( |\vec{q}|, |q_0|)}{\partial |\vec{q}|} \frac{\vec{q} \cdot \vec{p}}{| \vec{q}|} +  \frac{ \partial  f( |\vec{q}|, |q_0|)}{\partial |q_0|} \frac{q_0 \cdot p_0 }{| q_0|}\\
		& +  \frac{1}{2} \frac{ \partial^2  f( |\vec{q}|, |q_0|)}{\partial |\vec{q}|^2}\frac{(\vec{q} \cdot \vec{p})^2}{| \vec{q}|^2}  + \frac{1}{2} \frac{ \partial  f( |\vec{q}|, |q_0|)}{\partial |\vec{q}|} \left(\frac{\vec{p}\cdot \vec{p}}{| \vec{q}| } - \frac{(\vec{q} \cdot \vec{p})^2}{ |\vec{q}|^3} \right)\\
		& +\frac{1}{2}  \frac{ \partial^2  f( |\vec{q}|, |q_0|)}{\partial |q_0|^2} p_0^2 + \frac{ \partial^2  f( |\vec{q}|, |q_0|)}{\partial |q_0| \partial |\vec{q}|} \frac{\vec{p} \cdot \vec{q}}{| \vec{q}|}  \frac{q_0 \cdot p_0 }{| q_0|}+\mathcal{O}( p^3 ).
	\end{split}
\end{equation}
We can then apply this expansion to $\mathcal {G}_{\chi_{c} \chi_{d}}(p+q)$. As a result, the propagator and its derivatives depend on the loop momentum only. The tensor structures related to the external momenta can then be simplified by applying \eqref{intids1} and the evaluation of the traces proceeds along the same lines as the computation of the tadpole diagrams described in the previous subsection.


\end{document}